# Triplet Excitons Reconcile Charge Generation and Recombination in Low-Offset Organic Solar Cells: Efficiency Limits in a 5-State Framework

Jonathan L. Langentepe-Kong[a], Manasi Pranav[b], Safa Shoaee[ab,*], Dieter Neher[b,†]

[a]Heterostructure Semiconductor Physics, Paul-Drude-Institut für Festkörperelektronik (PDI), Leibniz-Institut im Forschungsverbund Berlin e.V., Hausvogteiplatz 5-7, 10117 Berlin, Germany
[b]Institute of Physics and Astronomy, University of Potsdam, Karl-Liebknecht Straße 24/25, 14476 Potsdam, Germany
Corresponding authors: *shoaee@pdi-berlin.de, †neher@uni-potsdam.de

**Abstract**
The power conversion efficiency of organic solar cells has recently improved beyond 20%. The active layers of these devices comprise of at least two organic semiconductors, forming a type II heterojunction. Hereby, the device performance is determined by the kinetic interplay of various species, including localized excitons, charge transfer states as well as charge-separated states. However, a model which describes all relevant photovoltaic measures has yet to be developed. Herein, we present a comprehensive 5-state rate model which includes both singlet and triplet charge transfer states and takes into account the formation, re-splitting and decay of the local triplet state, parametric in the respective energy offset. We show that this model not only describes key device properties such as charge generation efficiency, photoluminescence, electroluminescence and Langevin reduction factor simultaneously but also elucidate how these vary across material combinations based on the D:A interfacial energy offset alone. We find that the electroluminescence and Langevin reduction factor depend strongly on the triplet properties and that the triplet decay becomes the dominant charge recombination pathway for systems with moderate offset, in full agreement with previous experimental results. Validation against literature data demonstrates the model's ability to predict the device efficiency accurately. Subsequently, we identify material combinations with singlet exciton to charge transfer state energetic offset of roughly 150meV as particularly promising. Our model explains further why recent certified efficiency records for binary blends remain at ca. 20% if no further means to improve photon and charge carrier harvesting are taken.

## 1. Introduction

The rapid progress of organic solar cells (OSCs) in recent years, following the advent of non-fullerene acceptors (NFAs), has brought them to the forefront of photovoltaic research with the power conversion efficiency (PCE) having surpassed the 20% benchmark. Yet, significant voltage and fill factor (FF) loss in OSCs compared to inorganic devices remains a major challenge, even in high-performing systems.

It is widely accepted that free charge generation in NFA-based OSCs, following light absorption into bound singlet excitons (LE), is mediated by interfacial charge transfer (CT) states, constituting an electron-hole pair extending over the donor:acceptor interface which is formed after hole transfer from the acceptor to the donor [1]. The formation of these intermediary CT states relies heavily on the energetic driving force between LE and CT, often estimated as the highest occupied molecular orbital (HOMO) energy offset $\Delta_{HOMO}$ between donor and acceptor. At the same time, this driving force introduces a significant energy loss, seen as the reduction of the open circuit voltage, $V_{oc}$, to values usually below 1V, much lower than the optical bandgaps of the organic materials would imply. Reducing said offset to minimize the voltage loss in turn comes at the expense of the free charge generation and leads to a reduction of the short circuit current density, $J_{sc}$, and worsening of the FF [2].

Recently, studies have shown material combinations with minimal driving force that still achieve high performance [3] [4] [5] - suggesting that current generation remains uncompromised. Reported energy levels, however, show a large variation based, in part, on the different measurement methods, fabrication techniques and materials used [6]. We recently highlighted the critical role of LE dissociation (probed by photoluminescence) in charge generation using a two-state-model, wherein even at high CT dissociation efficiencies, a $\Delta_{HOMO}$ of at least 0.3eV is needed to give rise to efficient free charge generation that is not primarily compromised by LE decay [7]. Previous estimates of the minimum offset were set at roughly 0.5eV [8]. Convincing fits to both the internal quantum efficiency (IQE) as well as the non-radiative voltage loss (via electroluminescence) were presented by Classen et al., emphasizing the importance of long exciton lifetimes to counteract the slow exciton dissociation resulting from small driving forces [9]. Price et al. have shown the effectiveness of such kinetic models at recreating time-resolved measurements, such as transient absorption spectroscopy (TAS) [10]. Sandberg et al. expand this framework to predict PCE limits and uncover kinetic requirements to reach these [11]. However, a holistic understanding of OSC device performance remains incomplete without consideration of triplet states and especially their reseparation [12] [13], which are believed to impact charge recombination detrimentally - affecting the non-radiative voltage loss and the fill factor.

In this work, we aim to develop a comprehensive model capable of simultaneously reproducing multiple key metrics, including the internal generation efficiency (IGE), photoluminescence quantum yield (PLQY), electroluminescence quantum yield (ELQY) and the Langevin reduction factor (γ). As such, we show how the simple two-state model, based solely on LE and CT states, inadequately describes the entire solar cell physics. We resolve these shortcomings by introducing triplet states – which are undoubtedly present in organic material systems [14] [15] [16] [17] [18]. Our analysis reveals that the interplay between the different states governs the device physics and is vital to understand for improving the efficiency of organic photovoltaics.

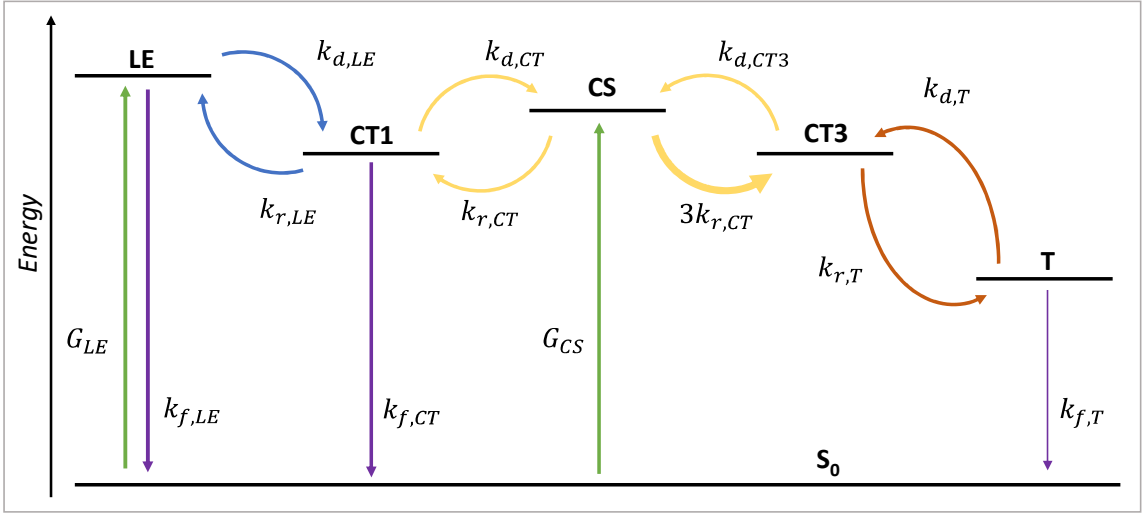

**Figure 1: Schematic energy level diagram including relevant kinetics for a 5-state-model of organic solar cells.** It is made up of localized singlet and triplet excitons (LE and T), the singlet and triplet charge transfer states (CT1 and CT3) as well as the separated charges (CS). Generation rates into LE and CS ($G_{LE}$ and $G_{CS}$) serve as input, while decay from LE, CT1 and T ($k_{f,LE}$, $k_{f,CT}$ and $k_{f,T}$) serve as the loss channels. Competing processes on CT1 include dissociation and reformation of the LE ($k_{d,LE}$ and $k_{r,LE}$) and of the CS ($k_{d,CT}$ and $k_{r,CT}$). Analogously, the triplet side includes dissociation and formation of the T ($k_{d,T}$ and $k_{r,T}$) and of the CT3 ($k_{d,CT3}$ and $3k_{r,CT}$, with a triplet degeneracy factor of 3).

## 2. Theory and Model
### 2.1 Modelling organic solar cell kinetics

In typical solar cell operation, excitons are photogenerated through light absorption ($G_{LE}$) and assumed to occupy a defined energy level, the acceptor singlet exciton energy. These either decay to the ground state $S_0$ (with a lifetime of $\tau_{LE} = k_{f,LE}^{-1}$) or dissociate ($k_{d,LE}$) into singlet CT states (CT1). These can either back transfer to the LE ($k_{r,LE}$), decay to $S_0$ ($\tau_{CT} = k_{f,CT}^{-1}$) or ideally dissociate further ($k_{d,CT}$) into free charges (CS). Direct generation of CS is also possible through electro-injection ($G_{CS}$), after which they non-geminately recombine ($k_{r,CT}$) to populate either the triplet CT state (CT3) or the CT1 in a 3:1 ratio, according to quantum mechanics. CT3 can again dissociate ($k_{d,CT3}$) into CS or form ($k_{r,T}$) localized triplet excitons (T). Lastly, these triplets either decay to $S_0$ ($\tau_T = k_{f,T}^{-1}$) or re-dissociate ($k_{d,T}$) to CT3.

Importantly, the direct decay of CS or CT3 to the ground state is neglected, so is the exchange between the different CT or exciton populations due to intersystem crossing being too slow to compete with the other pathways; $k_{ISC}$ has been shown to be orders of magnitude below singlet (LE and CT1) decay and dissociation rate coefficients [13] [14]. Extraction of free charges is also not considered explicitly, as described below. The two CT dissociation constants ($k_{d,CT}$, $k_{d,CT3}$) are assumed to be equal, due to their degeneracy $E_{CT1} \cong E_{CT3}$, but remain distinguishable in the equations.

The model described here is visualized in Figure 1 and gives rise to the following system of rate equations:

$$\frac{dn_{LE}}{dt} = G_{LE} - n_{LE}(k_{d,LE} + k_{f,LE}) + n_{CT1} \cdot k_{r,LE} \quad (1a)$$

$$\frac{dn_{CT1}}{dt} = n_{LE} \cdot k_{d,LE} - n_{CT1}(k_{r,LE} + k_{d,CT} + k_{f,CT}) + n_{CS}^2 \cdot k_{r,CT} \quad (1b)$$

$$\frac{dn_{CS}}{dt} = G_{CS} + n_{CT1} \cdot k_{d,CT} - n_{CS}^2 \cdot 4k_{r,CT} + n_{CT3} \cdot k_{d,CT3} \quad (1c)$$

$$\frac{dn_{CT3}}{dt} = n_{CS}^2 \cdot 3k_{r,CT} - n_{CT3}(k_{d,CT3} + k_{r,T}) + n_T \cdot k_{d,T} \quad (1d)$$

$$\frac{dn_T}{dt} = n_{CT3} \cdot k_{r,T} - n_T(k_{d,T} + k_{f,T}) \quad (1e)$$

In steady-state conditions, where $\frac{dn_x}{dt} = 0$, this system of equations is solved for the population densities $n_x$, as detailed in the Supplementary Information, Supplementary Note 2.5.

### 2.2 Device Metrics
**Internal Generation Efficiency**

The efficiencies associated with the LE and CT dissociation processes ($\eta_{d,LE}$ and $\eta_{d,CT}$) are given by considering the competition between the different processes from each state:

$$\eta_{d,LE} = \frac{k_{d,LE}}{k_{d,LE} + k_{f,LE}} \quad (2a)$$

$$\eta_{d,CT} = \frac{k_{d,CT}}{k_{r,LE}^{eff} + k_{d,CT} + k_{f,CT}} \quad (2b)$$

where $k_{r,LE}^{eff} = [1 - \eta_{d,LE}] \cdot k_{r,LE}$ is the effective back transfer rate coefficient to LE, reduced by the LE dissociation efficiency, cf. [11].

The free charge (photo)generation efficiency is the product of these two efficiencies (more in Supplementary Note 2.2), cf. [7]:

$$\eta_{CG} = \eta_{d,LE} \cdot \eta_{d,CT}$$
$$= \frac{k_{d,LE} \cdot k_{d,CT}}{(k_{d,LE} + k_{f,LE})(k_{d,CT} + k_{f,CT}) + k_{r,LE} \cdot k_{f,LE}} \quad (3)$$

and is used to fit the IGE - the ratio of the number of generated free charges to the number of absorbed photons - as one of the key metrics determining the efficiency of a solar cell. Note that the IGE is different from the internal quantum efficiency which also accounts for charge extraction, which we do not consider here.

**Photo- and Electroluminescence Quantum Yield**

We define the *PL decay efficiency* at $V_{oc}$ as the efficiency of LE decay when utilizing only photoexcitation:

$$\eta_{PL} = \frac{n_{LE}(G_{CS} = 0) \cdot k_{f,LE}}{G_{LE}} \quad (4)$$

Mirroring this, the *EL decay efficiency* is defined as the efficiency of LE decay when utilizing only electro-injection:

$$\eta_{EL} = \frac{n_{LE}(G_{LE} = 0) \cdot k_{f,LE}}{G_{CS}} \quad (5)$$

These decay efficiencies represent the fraction of total excitations (either $G_{LE}$ or $G_{CS}$) that decay through the LE state. We will use these efficiencies to fit the experimental PLQY and ELQY data, which provide valuable insight into charge recombination and voltage loss mechanisms [19]. However, the decay efficiency cannot be measured directly because only a fraction of the total decay is *radiative* and measurable as luminescence while the remainder decays to the ground state *non-radiatively* [20].

In order to compare the experimental data to the calculated result from Equation (4), the measured PLQY of the blend is normalized by that of the acceptor. Following absorption and Förster resonance energy transfer (FRET), we assume all excitons reside on the lower-bandgap material, even if blended with another material [8] [11]. Consequently, recombination proceeds via the acceptor LE, rather than the donor, and the radiative fraction of exciton decay in donor-acceptor blends (D:A) is identical to that of the neat acceptor. We have validated this in a previous work by modelling the blend PLQY using the neat PLQY for various blends (in the absence of LE-CT hybridization) [7].

Optical outcoupling issues that additionally reduce the measured luminescence signal - such as reabsorption in the device, cavity effects arising from reflective interfaces, $2\pi$-radiation, etc. - are also corrected for using the same approach. Experimentally, the neat acceptor is dispersed in a polystyrene matrix (PS:A) with comparable thickness, to more closely resemble the morphology and aggregation of the acceptor in the D:A blend. Thus, the experimental PL decay efficiency is calculated as (Supplementary Note 2.2):

$$PLE = PLQY_{D:A}/PLQY_{PS:A} \quad (6)$$

Given identical sample architecture and measurement configuration, which ensures the same radiative fraction and outcoupling efficiency, the experimental EL decay efficiency is similarly calculated as:

$$ELE = ELQY_{D:A}/PLQY_{PS:A} \quad (7)$$

We acknowledge here that hybridization of the LE with the CT1 may occur in low-offset systems [21] [22]. This will potentially increase the radiative decay rate of CT1, meaning that the blend emission is no more given only be the radiative decay of LE. Detailed 3-level calculations as well as experiments however showed the contribution of the CT emission to the total emission remain small even in case of a small energy offset [23]. We also see no evidence of heightened CT emission in our experimental data, as visualized in Figure S4. The other extreme to be considered is a large LE-CT offset where due to fast LE dissociation and slow LE reformation, CT1 decay becomes a well visible if not dominating radiative decay channel [22]. Our model assumptions will be no more valid for such high offset systems but such systems are not in our interest as they generally exhibit very large voltage losses.

**Langevin Reduction Factor**
Recombination in disordered, low-mobility systems such as in OSCs is typically described as free charge encounters leading to CT states, as explained by Langevin theory [24]. The Langevin recombination coefficient in our model is thereby given as:

$$k_L = 4k_{r,CT} \quad (8)$$

where the 4 accounts for encounters to both singlet and triplet CT states. In the steady-state condition, the state densities reach equilibrium, whereby the generation into CS is balanced by the bimolecular recombination from CS. Therefore, when $G_{LE} = 0$,

$$R_2 = k_2 \cdot n_{CS}^2 = G_{CS} \quad (9)$$

See Equation S5.8 for a more extensive look. The Langevin reduction factor, an important metric to describe the underlying recombination mechanisms, is then defined as the ratio:

$$\gamma = \frac{k_2}{k_L} \quad (10)$$

which describes the suppression of charge recombination ($k_2$) relative to the Langevin limit ($k_L$). Increased charge generation efficiency (due to CT dissociation) as well as suppressed triplet exciton formation was previously shown as a means to reduce bimolecular recombination [25] [26]. Equation (15) shows how the reduction factor can be written in terms of the (singlet and triplet) CT re-dissociation after their formation from CS encounters.

**2.3 Transfer Rates According to Marcus Theory**
The famous Marcus transfer rate coefficient given in Ref. [27] reads

$$k(\lambda, \Delta E) = \frac{H_{ab}^2}{\hbar}\sqrt{\frac{\pi}{\lambda k_B T}} \exp\left\{-\frac{(\lambda + \Delta E)^2}{4\lambda k_B T}\right\} \quad (11)$$

with driving force $\Delta E$ (as the Gibbs free energy difference between the initial and final state), temperature T, Boltzmann constant $k_B$ and reduced Planck constant $\hbar$. The electronic coupling strength $H_{ab}$ and reorganization energy $\lambda_{LE}$ have been obtained from DFT calculations in [7]. Note that *singlet reorganization energy $\lambda_{LE}$* will be used as a shorthand for the reorganization energy of the charge transfer process between the LE and CT1 (and similarly: the *triplet reorganization energy $\lambda_T$* for the T-CT3 process). The total value consists of the inner and outer reorganization energy, where values of around 200 meV have been reported for the latter [28].

We model the exciton dissociation and (re)formation rate coefficients in the Marcus frame, as commonly employed [29] [30]:

$$k_{d,LE} = 0.1\, k(\lambda_{LE}, -\Delta E_{LE-C}) \quad (12a)$$
$$k_{r,LE} = k(\lambda_{LE}, \Delta E_{LE-CT}) \quad (12b)$$
$$k_{d,T} = k(\lambda_T, -\Delta E_{T-CT}) \quad (12c)$$
$$k_{r,T} = k(\lambda_T, \Delta E_{T-CT}) \quad (12d)$$

where the driving force for the LE to dissociate into CT states is provided by their energy difference and exciton splitting readily occurs for a positive offset, i.e. $\Delta E_{LE-C} = E_{LE} - E_{CT} > 0$. For triplet excitons, we similarly define the driving force as $\Delta E_{T-CT} = \Delta E_{LE} - \Delta E_{ST}$, with the singlet-triplet energy difference $\Delta E_{ST} = E_{LE} - E_T$. In organic, phase-separated bulk heterojunctions (BHJs), where most photoexcited NFA molecules lack a neighbouring donor molecule to undergo hole transfer, the limited diffusion of the exciton to the interface is considered via a 0.1 prefactor in the singlet exciton dissociation rate coefficient [31]. It can also be interpreted as a mismatch of the density of states at the interface (CT) and in the bulk (LE): g(CT)/g(LE) = 10, like in Ref. [9]. The triplet excitons, only formed at the interfaces from CT3 recombination, such that g(T) = g(CT3), lack this prefactor.

**Table 1: Parameters used for the steady state model.** Most values are fixed to typical, representative values. Some values are slightly variable (denoted by ≈) and will act as our fitting variables, within reasonable ranges. The driving force $\Delta E_{CT-LE}$ acts as the x-axis. There are 8 primary parameters defining the model, indicated **bold**. The CT1 reformation is treated as a free parameter in the model and can be set arbitrarily due to its balance with CT dissociation.

| Symbol | Description | Value | References |
|---|---|---|---|
| $G_{LE}, G_{CS}$ | Photogeneration and electro-injection rates | $1 \times 10^{28}$ m$^{-3}$s$^{-1}$ | |
| **$k_{f,LE}$** | Singlet exciton decay | $1 \times 10^9$ s$^{-1}$ | [1], [9], [33], [34], [35] |
| **$k_{f,CT}$** | CT1 decay | $1 \times 10^9$ s$^{-1}$ | [29], [25], [39] |
| **$k_{f,T}$** | Triplet decay | $\approx 1\text{-}10 \times 10^6$ s$^{-1}$ | [16] |
| **$k_{d,CT}, k_{d,CT3}$** | CT dissociation | $1 \times 10^{11}$ s$^{-1}$ | [14] |
| $k_{r,CT}$ | CT1 reformation | $2.5 \times 10^{-16}$ m$^3$s$^{-1}$ | Supplementary Material |
| $k_{d,LE}, k_{d,T}$ $k_{r,LE}, k_{r,T}$ | Exciton dissociation and exciton reformation | Marcus Rates ↴ | |
| **$H_{ab}$** | Coupling strength | 0.01eV | [7], [13], [32] |
| **$\lambda_{LE}$** **$\lambda_T$** | Singlet and triplet reorganization energy | $\approx 0.5$eV $\approx 0.2$eV | [7], [30] [29] |
| **$\Delta E_{ST}$** | Singlet-triplet offset | $\approx 0.3$eV | [41] |
| $\Delta E_{LE-CT}$ | Driving force | -0.4eV to 0.4eV | |

The aforementioned parameter values used in the model are condensed in Table 1. Efficient CT dissociation throughout is maintained by offset-independent CT dissociation rate coefficients, supported by the findings in Figure S7a, wherein inefficient CT dissociation fails to describe the experimental data. Figure S5 gives a more nuanced look on inefficient CT dissociation at low-offset.

Here, reported numbers of Y6 and its binary blend with PM6 serve as guidance. Nevertheless, even for these well-studied materials, some values vary significantly. For example, while values for the LE-CT1 electron coupling in PM6:Y6 from DFT calculations are rather consistent (around 10 meV), the prediction of the inner reorganization energy ranges quite significantly between 100 meV and 325meV [7] [13] [32]. Also, while most groups reported a LE decay time of neat Y6 of around 0.8-1.5 ns [9] [33] [34] [35], shorter PL decay times were occasionally reported, such as 250ps [36]. The situation is even more complicated for the CT1 decay times which will depend, among others, on the CT energy following the energy gap law [37] [38]. However, a decay time of ca. 1 ns seems to be reasonable value for D:A blends with not too small a $V_{oc}$ (and therefore CT energy) [25] [38] [39].

Studies reporting the reorganization energy of the triplet (T-CT3) process are quite scarce and do not seem to agree. $\lambda_T$ is reported to be similar to that of the singlet (LE-CT1) process in Ref. [13]; much higher for recombination to triplet excitons than for singlet exciton dissociation in Ref. [40]; or much lower, around 100meV, in Ref. [29]. We already note here, that our model also does not yield a definitive conclusion on $\lambda_T$ (see Sec. 4.2). Triplet lifetimes are reported to be on the order of tens of microseconds [16], but can be reduced to only a few microseconds or less by triplet-polaron annihilation [14]. Furthermore, the LE energy in NFAs was shown to be greatly reduced by their fluorinated end-group π-π stacking, resulting in $\Delta E_{ST}$ as low as 0.3eV; demonstrated for Y6 [41].

## 3. Analysis: Model Comparisons

The exact solutions to all models introduced here can be found in the Supplementary Information, Supplementary Notes 2.1-2.5.

### 3.1 The Singlet-Branch (the 2 State Model)

Since, in the absence of intersystem crossing, the free charge photogeneration efficiency $\eta_{CG}$ is solely determined by the LE and CT1 dissociation efficiencies (see Eq. 3), it is unaffected by the presence or kinetics of the CT3 or T states, and should therefore remain unchanged when these states are introduced. The PL efficiency $\eta_{PL}$, largely playing in the singlet branch as well, should similarly remain independent. Figure 2a visualizes the changes between the simplest 2-state model, encompassing only the LE and CT1 state, and the complete 5-state model as described above. Note, that we plot $\Delta E_{LE-C}$ here, mirrored to $\Delta E_{CT-LE}$ in our recent work [7].

In both models, these two metrics - $\eta_{CG}$ and $\eta_{PL}$ - can be very accurately fitted using the same set of parameters, with $\lambda_{LE} \approx$ 550meV- the reorganization energy for the LE to CT1 hole transfer. The IGE and PL trend is straightforward. A negative driving force (LE below CT1) prevents LE dissociation, whereby almost all excitons remain in the LE and decay from there. A positive driving force (LE above CT1) results in the LE dissociation depopulating the LE state in favour of the CT1 and reducing the LE decay. With efficient CT dissociation, the CT1 population is directly linked to free charge generation, Equation (S2.1). In the "transition region" in between, the LE reformation and dissociation compete - the balance being determined by the offset, Equation (26) - resulting in the sigmoidal shape in accordance with the experimental data.

Figure 2a also demonstrates how the charge generation is in direct competition with the LE decay, in agreement with Ref. [7], maintaining the relationship $\eta_{CG} + \eta_{PL} = 1$ throughout. Only in the case of very efficient LE dissociation does a slight deviation appear, stemming from losses from the CT decay due to high CT populations (cf. Fig. 5), resulting in $\eta_{CG} + \eta_{PL} < 1$. Interestingly, the transition region in the 5-state model shows $\eta_{CG} + \eta_{PL} > 1$. While the sum of these efficiencies exceeding unity might initially appear unphysical, it simply arises from free charge recombination back to CT1 (and subsequently LE), which was not considered in the 2-state model, thereby counting towards both $\eta_{CG}$ and $\eta_{PL}$. This behaviour is restricted to the transition region, where neither the LE reformation nor the LE dissociation dominates, so that $k_{r,LE}^{eff}$ is non-zero. Note that some transient experiments such as time-delayed collection field (TDCF) or transient absorption (TAS) allow probing of the initial state of free charge generation without contributions from CS recombination. Consequently, the results from these measurements can be described by the 2-state model.

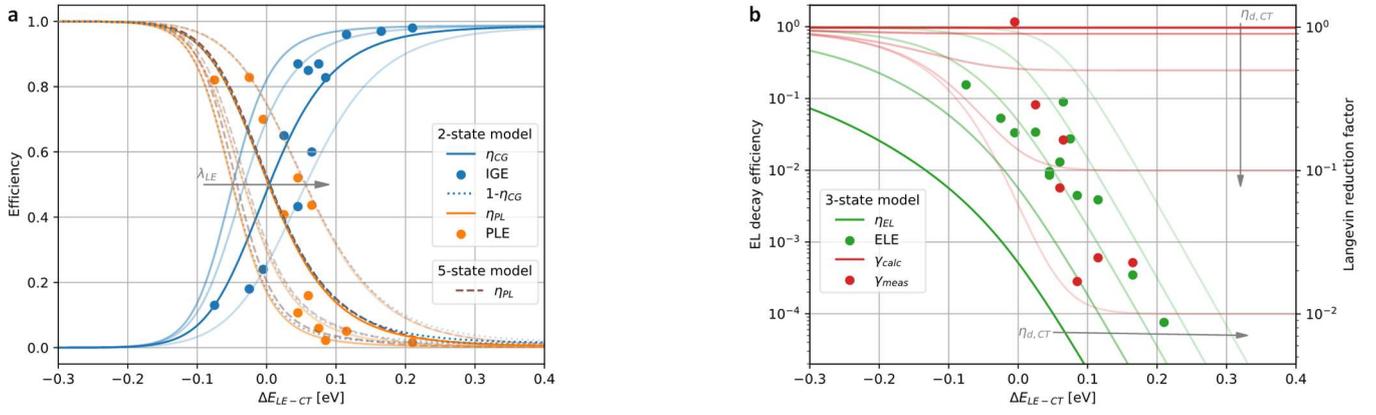

**Figure 2: Two states capture the IGE and PLQY accurately. Three states describe but overestimate ELE.** a) Free charge generation and PL decay efficiency, $\eta_{CG}$ and $\eta_{PL}$, in the 2-state model - shown for different singlet reorganization energies $\lambda_{LE}/eV = 0.35, 0.45, \mathbf{0.55}, 0.65$ as indicated by the arrow. It can be seen how the singlet decay competes with free charge generation with $\eta_{PL} = 1 - \eta_{CG}$, provided CT dissociation is fast, cf. *[7]*. The PL decay efficiency in the 5-state model is shown for comparison (dashed brown line). b) EL decay efficiency and Langevin reduction factor, $\eta_{EL}$ and $\gamma_{calc}$, in the 3-state model where free charges encounter to (re)form CT1 states – shown for increasing CT1 dissociation efficiencies, $\eta_{d,CT} = 0.01, 0.1, 0.5, 0.95, 0.99$ following the arrows' directions. The experimental data are overlaid as circles.

### 3.2 Free Charge Recombination (the 3 State Model)
We now turn to the prediction of the electroluminescence efficiency and the Langevin reduction factor, enabled by the CS state and its generation rate $G_{CS}$. Since we do not consider trap-assisted recombination, free charge recombination must proceed through the reformation to and the subsequent decay of the CT state (and to a lesser extent the LE state). We mention already here that neither $\eta_{EL}$ nor $\gamma$ depends on the value of the bimolecular encounter rate coefficient of free charges to CT1, $k_{r,CT}$, due to the balance of CS encounters with CT dissociation, c.f. [11]. A smaller (higher) $k_{r,CT}$ will only increase (decrease) $n_{CS}$, without affecting the CT or LE populations.

Figure 2b visualizes both $\eta_{EL}$ and $\gamma$ for different values of the CT1 dissociation efficiency $\eta_{d,CT}$. The 3-state model captures the overall dependence of ELE and $\gamma$ on $\Delta E_{LE-CT}$, both decreasing with offset, but fails to predict their absolute values. The lightest line corresponds to the values given in Table 1, with $\eta_{d,CT} = 0.99$ through $k_{d,CT} = 1 \cdot 10^{11} s^{-1}$ and $k_{f,CT} = 1 \cdot 10^{9} s^{-1}$. For this set of parameters, the ELE is overestimated by more than an order of magnitude, and the reduction of $\gamma$ happens roughly 0.1eV earlier than anticipated. The ELE values can be reconciled by assuming a dissociation efficiency less than unity, i.e. $\eta_{d,CT} \approx 0.5$. However, this is in contrast with the charge generation pathway, where efficient CT1 dissociation is necessary to reach an IGE close to unity as in the 2-state model (Fig. S2a).

Regarding the coefficient of free charge recombination, it is well established that $\gamma$ is determined by the capability of the CT state to reform free charges in competition with its decay to the ground state; a large reduction factor (close to unity) indicates inefficient CT resplitting and vice versa [42], as clearly shown in Figure 2b. While varying $\eta_{d,CT}$ values can reproduce certain individual $\gamma_{meas}$ data points, they do not provide a consistent description across the entire data set.

A systematic dependence of $\gamma$ on $\Delta E_{LE-C}$ is, indeed, expected from the energy gap law: Larger (more positive) $\Delta E_{LE-}$ implies a smaller CT energy and consequently higher $k_{f,CT}$, which would decrease $\eta_{d,CT}$ and correspondingly increase $\gamma$. However, the opposite trend is observed experimentally. A recent publication reported an increase of $k_{f,CT}$ for decreasing offset, which was attributed to an increasing coupling of the CT1 to a short lifetime LE [43]. Figure 2b indeed shows that the experimental $\gamma$ values are well reproduced by increasing the CT lifetime as the energy offset decreases (a reduction in $\eta_{d,CT}$), with the bold line representing $\Delta E_{LE-C} \approx 0$ and the lightest line corresponding to $\Delta E_{LE-C} \approx 0.15$. On the other hand, significantly increasing $k_{f,CT}$ to explain the large $\gamma_{meas}$ for our small offset systems decreases $\eta_{CG}$ and $\eta_{PL}$ well below experimental values, meaning that this approach is not appropriate (Fig. S2a).

Alternatively, $k_{d,CT}$ could be altered for a similar effect. For example, a recent work showed the Langevin reduction to be strongly related to the quadrupole moment of the NFA due to band bending [44]. As band bending also reduces the LE-CT splitting [8], a correlation between $\gamma$ and $\Delta E_{LE-C}$ via a systematic change of $k_{d,CT}$ may be plausible. However, a reduction of $\gamma$ with decreasing offset is expected from these considerations in contrast to the experimental findings. Also, as in the previous approach, a consistent explanation of all four metrics is needed here but not possible due to the dependence on the CT dissociation efficiency. Notably, $\eta_{EL}$ (and $\eta_{PL}$) is currently independent of $k_{d,CT}$ because all dissociated charges reform to CT, thereby not changing the LE-CT balance.

To conclude this section, we find that both the 2-state and 3-state models, which consider only singlet local excitons and CT states, adequately reproduce the experimental IGE and PLE data. However, they completely fail to capture the underlying recombination pathway of the ELE or $\gamma$.

### 3.3 The Triplet-CT (the 4 State Model)
It is unsurprising that the 3-state model is inadequate at describing recombination in OSCs, as only one out of three free charge encounters (i.e. spin uncorrelated) events form singlet CT states; the remaining 75% populate the triplet CT state [45]. This CT3 is characterized by both a re-dissociation to CS (equal to that of CT1: $k_{d,CT3} = k_{d,CT}$) and, in a first order approximation, a constant decay rate coefficient $k_{f,CT3}$. With this, the CT3 is added as an additional (non-radiative) decay channel.

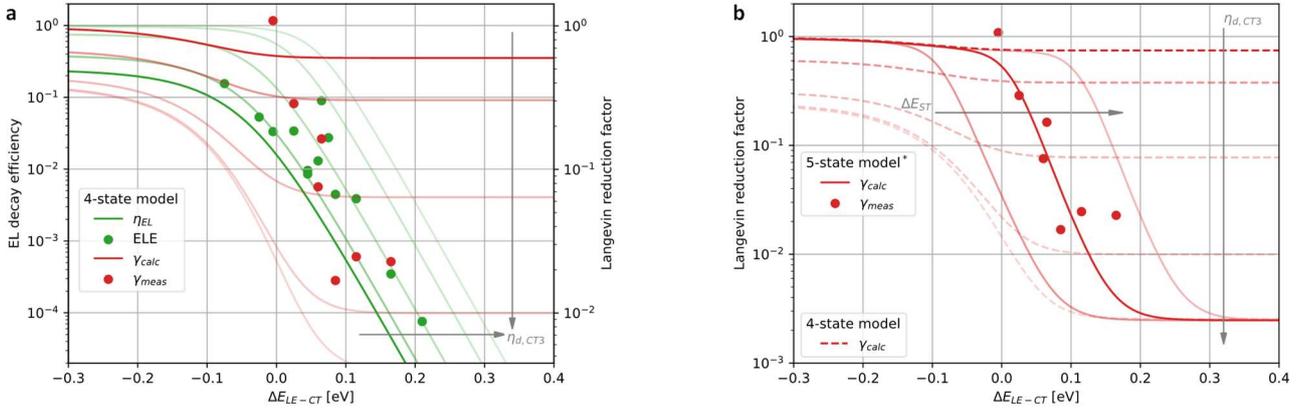

Figure 3: Including localized triplets unlocks simultaneous description of recombination via ELE and $\gamma$. a) The Langevin reduction factor $\gamma_{calc}$ and $\eta_{EL}$ calculated in the 4-state model with increasing CT3 dissociation efficiencies $\eta_{d,CT3}[\%] = 1, 50, 90, 99, 99.99$ as indicated by the arrows. The same curves are shown dotted in the next panel: b) The Langevin reduction factor calculated in the 5-state model (solid lines) for increasing singlet-triplet offset $\Delta E_{ST}[meV] = 200, 300, 400 meV$ as indicated by the arrow. The triplet exciton formation and dissociation are modelled as Marcus transfer rates (*with a constant inverted regime), allowing the model to reach higher $\eta_{d,CT}$ values at increasing offset.

Under these assumptions, the dissociation efficiency of the CT3 state becomes the new free parameter in the 4-state model used to fit the data, and is defined as:

$$\eta_{d,CT3} = \frac{k_{d,CT3}}{k_{d,CT3} + k_{f,CT3}} \qquad (13)$$

The impact of $\eta_{d,CT3}$ on the LE decay efficiencies is similar to the impact of $\eta_{d,CT}$ (Fig. S2). It is evident that the two variables are connected: with complete CT3 dissociation, and thereby no CT3 decay, all free charges eventually reform through CT1. However, for smaller $\eta_{d,CT3}$, charges are lost through CT3 and do not re-dissociate into CS, reducing the reformation to LE considerably, just as a larger $k_{f,CT}$ does. Interestingly, the case $\eta_{d,CT3} = 0$ (darkest lines in Fig. 3a) does not predict $\gamma$ to be one. The reason is that 25% of charge encounters populate CT1 which, due to their rapid resplitting, reduces recombination.

Figure 3a shows how at higher CT3 dissociation efficiencies, charges undergo multiple cycles of recombination and re-dissociation, leading to a reduction in $\gamma$. The impact of $\eta_{d,CT3}$ on $\gamma$ is distinct, in that it affects the entire driving force range, whereas $\eta_{d,CT}$ primarily acts on the positive offset, where LE dissociation is efficient (Fig. 2a). Still, the 4-state model, with only a constant CT3 decay rate coefficient, remains inadequate at recreating the Langevin reduction factor. A comparison between the experimental data and the different $\eta_{d,CT3}$ cases in the 4-state model (Fig. 3a) provides insight into how incorporating a *variable* CT3 dissociation efficiency can improve agreement with experimental observations. This model suggests a sharp transition in $\eta_{d,CT3}$ at low energy offset. Specifically, a low $\eta_{d,CT3}$ (darkest line) for $\Delta E_{LE-C} \lesssim 0$ eV, which quickly transitions to a high $\eta_{d,CT3}$ (lightest line) for $\Delta E_{LE-CT} \gtrsim 0.1$ eV, would capture the behaviour of the Langevin reduction factor. Since the CT3 dissociation rate coefficient is assumed to be constant, in line with that of CT1, we require a variable CT3 loss pathway to establish such a new transition.

### 3.4 The Triplet Excitons (the 5 State Model)

Here, it is natural to adopt a Marcus framework, Equation (11), to populate localized triplet excitons (T) through the recombination of CT3 states. The direct decay from CT3 to $S_0$ is replaced by triplet formation, $k_{f,CT3} \to k_{r,T}$, such that the spin-forbidden CT3 decay is removed from the model. Re-dissociation $k_{d,T}$ of the triplet to CT3 [12] is also modelled according to Marcus theory, while the triplet decay $k_{f,T}$ is treated as a constant value. The dissociation efficiencies of the triplet exciton and CT3 are obtained analogously to Equations (2a) and (2b):

$$\eta_{d,T} = \frac{k_{d,T}}{k_{d,T} + k_{f,T}} \qquad (14a)$$

$$\eta_{d,CT3} = \frac{k_{d,CT3}}{k_{d,CT3} + k_{r,T}^{eff}} \qquad (14b)$$

where $k_{r,T}^{eff} = [1 - \eta_{d,T}] \cdot k_{r,T}$. In contrast to Equation (13), $\eta_{d,CT3}$ is no longer a constant and allows for the new, sharp transition in $\gamma$, as described in the previous section. This transition is manifested by the competition of the formation and dissociation of the triplet, mirroring the LE competition already seen in the 2-state model. This LE competition already affects the Langevin reduction (Figs. 2a and 3a), creating an early transition starting around $\Delta E_{LE-CT} \approx -0.1$ eV. The new transition stemming from the competition on T is expected to occur at a different energy $\Delta E_{LE-} \approx +0.1$ eV to follow the experimental data - a shift of some hundreds of meV.

In this context, incorporating localized triplet excitons naturally satisfies both conditions:
1. The energy shift relative to the singlet competition given by the singlet-triplet energy difference $\Delta E_{ST}$
2. The variable rate/competition through the offset dependent triplet formation and dissociation rates via Marcus theory

For small offsets, the CT energy lies close to that of the LE and is therefore much higher than that of the triplets. Thus, formation of triplets (upon recombination of CT3) is highly efficient, while their re-dissociation is hindered by an energy barrier (of roughly $\Delta E_{ST}$). In this picture, the triplets act solely as a dead-end recombination channel, an equivalent picture to the 4-state model with $\eta_{d,CT3} = 0$ (Fig. 3a, dark line) where all CT3 recombine to $S_0$; either directly in the 4-state model or via triplets in the 5-state model. As the CT energy decreases (higher $\Delta E_{LE-CT}$), the triplet re-dissociation rate coefficient increases to the point where the triplet decay channel is no longer viable ($\eta_{d,T} = 1$). At this point, CT3 states fully dissociate into CS, with $\eta_{d,CT3} = 1$ (Fig. 3a, lightest line). The model sharply transitions between these two extremes.

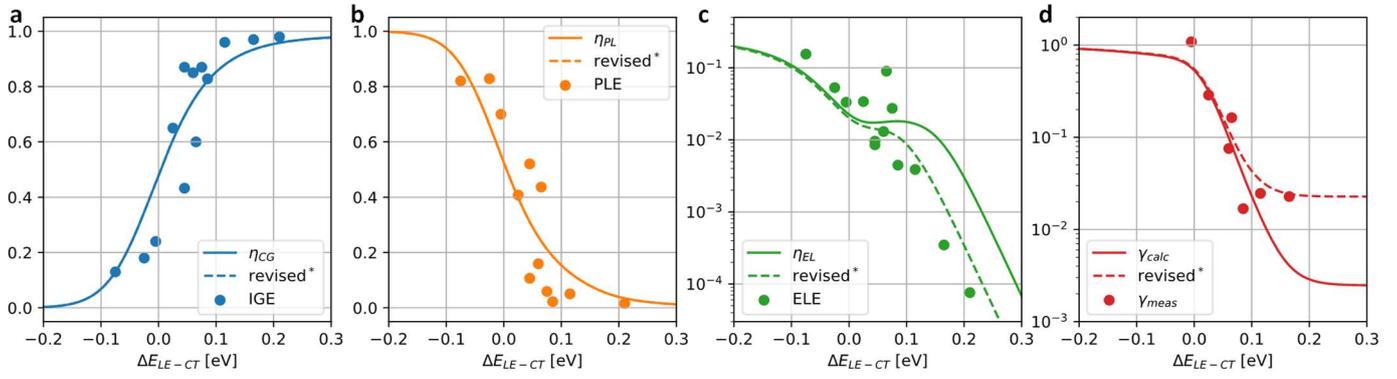

**Figure 4: CT state revisions refine key metrics, improving on the simple model.** The a) free charge generation efficiency, b) PL decay efficiency (Eq. 4), c) EL decay efficiency (Eq. 5) and d) Langevin reduction factor. All calculated in the simple 5-state model (solid lines) as described in Figure 1 and in the *revised 5-state model (dashed lines) where "deeper CT states" are incorporated, i.e. two different CT1 decay coefficients $k_{f,CT} = 10^9 s^{-1}$ and $10^{10} s^{-1}$ are used for the generation and recombination pathways, respectively.

This directly correlates with $\gamma$ through the following equation (derived from Equation S6.5, Supplementary Information):

$$\gamma = \left[\frac{1}{4}(1 - \eta_{d,CT}) + \frac{3}{4}(1 - \eta_{d,CT3})\right] \quad (15)$$

which expresses the suppression of charge recombination relative to the Langevin limit ($\gamma$) in terms of the re-dissociation efficiency of singlet and triplet CT states, $\eta_{dCT}$ and $\eta_{dCT3}$. Note that $\eta_{dCT3}$ and therefore $\gamma$ are heavily influenced by triplet kinetics through Equation (14): higher triplet dissociation (or suppressed triplet formation) correlates with reduced recombination, c.f. Ref. [25].

Figure 3b showcases the Langevin reduction factor, where a singlet-triplet offset of 300meV – a reasonable estimate for state-of-the-art NFAs [41] - creates a transition aligning well with the data. The simple Marcus rate coefficients (and thus $\gamma$) exhibit a turning point, as shown in Figure S3a (red dashed lines). While this behaviour predominantly lies outside of the experimental data range, it is rectified by setting the inverted Marcus region to a constant value (as in Fig. 3b). The extension to the Marcus-Levich-Jortner model, e.g. Ref. [30], would provide a more physically accurate description in this region. Nonetheless, for the range of offsets considered here and ignoring the temperature dependence, the constant inverted regime serves as a good approximation, splendidly reproducing the Langevin reduction factor along with the other metrics, as demonstrated in Figure 4.

### 3.5 The EL Bump
As previously mentioned, the IGE is unaffected by alterations in the triplet branch, which largely holds for the PL decay efficiency as well. Additionally, revising the rate coefficients in the inverted Marcus regime does not have a noticeable impact on the metrics other than $\gamma$.

However, the development of a "bump" in $\eta_{EL}$ is observed around $\Delta E_{LE-CT} \approx 0.1$eV, which does not align with the data (Fig. 4c, solid line). The offset dependent triplet formation $k_{r,T}$ was introduced to allow the model to pass through several different $\eta_{d,CT3}$ values, providing a much better approximation of the measured $\gamma$. However, Figure 3a also shows a change of the $\eta_{EL}$ with $\eta_{d,CT3}$: a lower (higher) CT3 dissociation efficiency implying smaller (larger) electro-luminescence, since 75% of injected charges form into CT3 states, significantly limiting the formation of LE. As previously mentioned, $\eta_{d,CT3}$ now increases with offset, such that $\eta_{EL}$ changes accordingly, and quickly jumps to a higher value as well, see Figure S3b. The model starts overestimating the ELE for positive offset. In the following, we discuss various approaches to reduce the significance of this bump and improve the model fit.

**Additional CT3 Decay**
It could be mitigated, if an upper limit to CS reformation from CT3 existed, realised through an additional decay rate from CT3 directly to the ground state, with coefficient $k_{f,CT3}$, causing some charges to be permanently lost through CT3 (in addition to T) and unable to form CT1 states. Figure S6d demonstrates how setting $k_{f,CT3} = 10^{10} s^{-1}$ almost eliminates the EL bump, which raises the question of whether such a strong decay from CT3 is feasible. With this coefficient, however, the Langevin reduction factor loses its transition enabled by triplet formation, because $\eta_{d,CT3} \ll 1$. A rate coefficient of $k_{f,CT3} = 10^9 s^{-1}$ is the highest CT3 decay possible before compromising on the fit to $\gamma$ but unable to completely remove the bump.

**Faster CT1 Decay**
A similar effect can be achieved by increasing the decay of CT1, as it would prevent more charges from reaching the LE from electro-injection (Fig. S6e). However, as part of the singlet branch, the alteration of CT1 heavily influences the generation efficiency as well, causing the upper limit $\eta_{CG,max} = \frac{k_{d,CT}}{k_{d,CT}+k_{f,CT}}$ to reduce (which was previously very close to unity due to $k_{f,CT} = 10^{-2} k_{d,CT}$). Figure S6 and Figure S7 further illustrate the impact of the various model parameters on the device metrics. The complex interdependencies make resolving the EL bump without affecting the other metrics difficult in the current model.

**CT Energetic Disorder**
A way out of this problem is to assign two different CT1 decay times to the generation and recombination routes. This can be rationalized by considering inhomogeneous energetic disorder, which broadens the density of states (DOS) of the (electronically and vibrationally relaxed) CT states. Because of this, CT1 states formed by exciton dissociation might be of higher energy than those reformed upon free charge encounter.

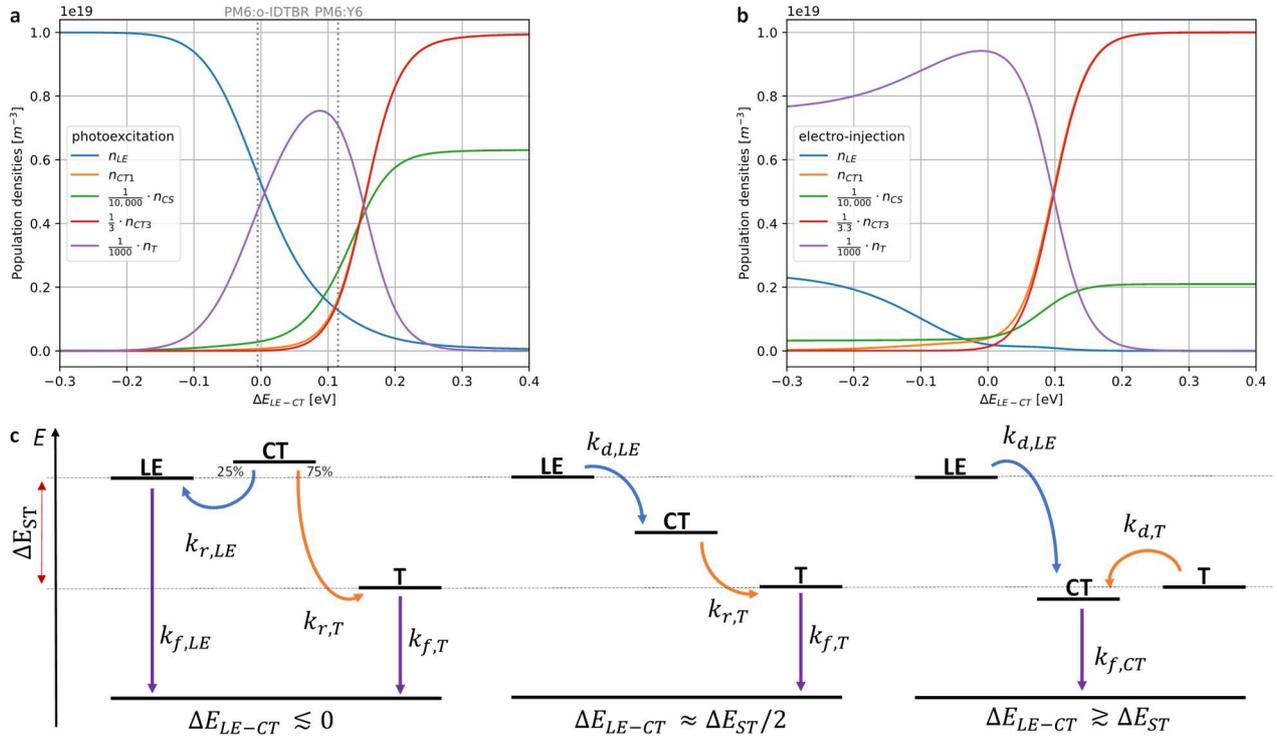

**Figure 5: Revealing population dynamics & non-radiative loss channels.** Steady-state population densities under photoexcitation (a) or electro-injection (b), modelled in the revised 5-state model; scaled to fit as indicated in the legends. Vertical lines show PM6:Y6 and PM6:o-IDTBR on the energy axis. c) Simplified state model diagramme, where CT stands for the degenerate CT1 and CT3, showing only the dominant processes. The triplet population and decay channel become highest for moderate offsets, while high-offset systems are dominated by CT decay.

Exciton dissociation is expected to occupy states higher in the CT1 DOS, especially for positive $\Delta E_{LE-}$ where the center of the LE DOS is at higher energies than that of CT1. In addition, photogenerated excitons have fairly short lifetimes, especially in D:A blends, rendering their equilibrium into states deep in the LE DOS less likely. Finally, recent work found the LE DOS to be on the order of 30-40 meV [46], which is much smaller than that of the charge transporting DOS from which the CT DOS is derived [47] [48]. Note that this concept of energetic disorder is very different from the model of hot CT formation and splitting [49] [50]. Here, LE dissociation populates vibronically excited CT states which are more delocalized and split more easily.

It has been concluded that steady-state charge recombination mainly proceeds among fully equilibrated charges [51] [52], at least at or near $V_{OC}$. This in turn will preferentially populate states deep in the CT1 DOS. According to the energy gap law, such "deeper" states exhibit a higher decay to the ground state, rendering their resplitting into CS more difficult and thus increasing $\gamma$. On the other hand, states at higher energy are more likely to undergo repeated recombination and separation events, compared to the deeper states [53]. The important role such states on recombination was proven in a recent study on a wide selection of D:A blends where a broader DOS led to a larger $\gamma$ [54], whereby $k_2$ being sensitive to energetic disorder [47] is easily understood.

This allows us to use two distinct CT1 decay rate coefficients, $k_{f,CT} = 10^9 s^{-1}$ and $10^{10} s^{-1}$ for photoexcitation and electro-injection, respectively. This is only a slight modification of the model, as all state populations equations can be expressed in the form $n_x = aG_{LE} + bG_{CS}$ to separate the charge generation and recombination pathways (Supplementary Note 2.5). Figure 4a-d (dashed lines) visualize the reduction of the EL bump in this revised model without compromising the other device metrics, especially $\eta_{CG}$ or $\gamma$.

## 4. Results of the 5-state model

It is now well-established that state models based solely on the singlet branch cannot simultaneously describe all four metrics as a function of energy offset at the D:A heterojunction. Omitting triplets from these models leads to an incomplete representation of the solar cell energetics and kinetics, especially free charge recombination as seen through the Langevin reduction. By contrast, our 5-state model accurately captures the device characteristics when utilizing experimentally observed energy levels, reorganization energies, dissociation rate coefficients and lifetimes, as summarized in Table 1 (note that the inclusion of "deeper CT states" from the previous section is not addressed in the table).

It is important to recognize that fitting an entire data set – comprising many materials and combinations – with a single set of parameters presents inherent challenges. Naturally, we do not expect one fit to capture the complexity of all material systems. This is evident in Figure S7d, where two different $H_{ab}$ values, 10meV and 5meV, fit the IGE and PLE data better than a single value; and similarly for $\lambda_{LE}$ in Figure S7c.

### 4.1 Competition between Charge Generation and Exciton Decay

Figures 2 and S1 illustrate how charge generation competes with singlet exciton decay, with $\eta_{CG} + \eta_{PL} \approx 1$ as discussed earlier. This competition arises from rapid CT dissociation relative to CT decay. Several factors, such as $\lambda_{LE}$, $H_{ab}$ and $k_{f,LE}$ (Fig. S7) simultaneously influence the shape of both curves, maintaining the competition. However, as shown in Figure S7a and Figure S6e relaxing the condition $k_{d,CT} \gg k_{f,CT}$ decouples the PLE and IGE. Interestingly, the triplet kinetics can also affect the coupling, e.g. by increasing the CT1 reformation through enhanced triplet dissociation, Figures S6a&b.

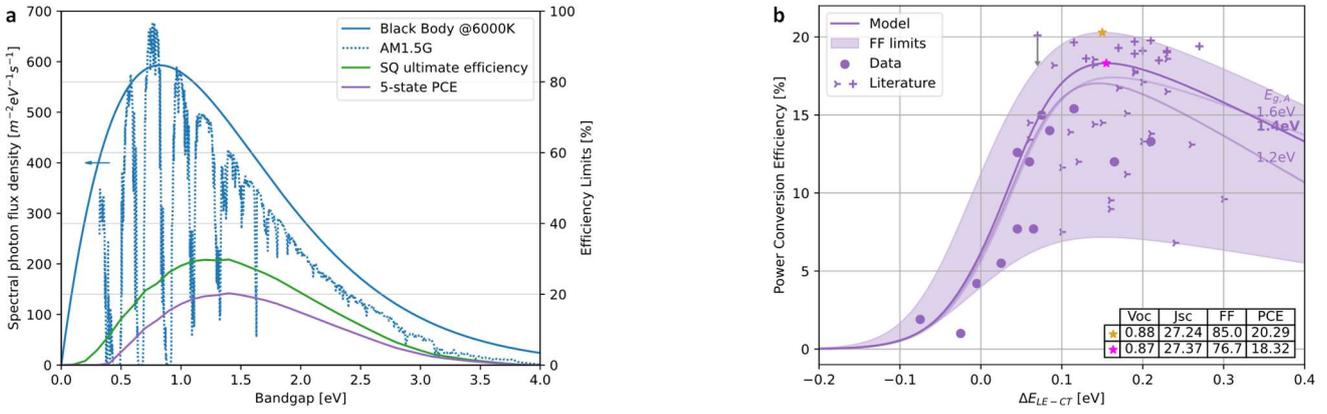

**Figure 6: Model calculations closely match experimental PCEs and reveal efficiency limits of OSCs.** a) Black body spectrum of the sun at $T_S = 6000K$ (solid blue line) and the measured AM1.5G solar spectrum (dotted blue line). Shockley-Queisser (SQ) ultimate efficiency (green) and the efficiency limit in the 5-state model (purple) are shown. b) Change of device efficiency with energy offset $\Delta E_{LE-C}$ for an optical bandgap of $E_g = 1.4eV$ (1.2eV and 1.6eV in lighter), an absorption of 90% above the bandgap and FF according to Eq. 18 (solid line) as well as between 85% and 30% (shaded area) with the maxima marked (stars). Literature values (from Tab. S2 and S3) are shown to complement our data (triangles and pluses), plotted for $\Delta_{HOMO}$ due to lacking $\Delta E_{LE-C}$ data. An arrow indicates where a particular outlier falls without the enhanced surface roughness that boosts its light-harvesting (LBL, 20.1%), corresponding to the BHJ of lower efficiency (18.4%).

### 4.2 The Role of Triplet Lifetime

Using a lifetime of 1μs, the triplet parameters are fitted to be $\lambda_T = 250meV$ and $\Delta E_{ST} = 300meV$ (Fig. 3b). The exact lifetime value matters little for the fit as the other parameters can be adjusted to restore it. The increase (decrease) in the lifetime is e.g. compensated by increasing (decreasing) the singlet-triplet offset, so that different combinations of triplet parameters can yield very similar results: $(k_{f,T}, \Delta E_{ST}) = (10^6 s^{-1}, 300meV)$, $(10^5 s^{-1}, 350meV)$ and $(10^7 s^{-1}, 250meV)$. Figures S6a-c reveal how each of these parameters affects the metrics independently, e.g. showing the minimal impact of the triplet reorganization energy, which leads to high uncertainty of this parameter as well as the triplet lifetime. We note that introducing diffusion limitation on triplets, i.e. $g(CT3)/g(T) > 1$, decreases their dissociation as well and is another parameter able to compensate changes in $k_{f,T}$ or $\Delta E_{ST}$.

### 4.3 Triplet Population and Non-Radiative Losses

Figure 5a gives further insight into the role of the triplet lifetime. The highest triplet population (under illumination) is found in systems with moderate offset, $\Delta E_{LE-CT} \approx 0.1eV$, due high LE dissociation and simultaneously low T dissociation at this offset (shown in the middle panel of Fig. 5c). The triplet population almost reaches $n_T^{max} = G_{LE}/k_{f,T} = 10^{22} m^{-3}$, three orders of magnitude higher than that of the singlet population limit $n_{LE}^{max} = G_{LE}/k_{f,LE} = 10^{19} m^{-3}$, stemming from the lifetimes being $\tau_T = 1000 \cdot \tau_{LE}$.

From the triplet state density in Figure 5a (purple), the triplet decay efficiency under illumination can be calculated in accordance with the PL decay efficiency from Equation (4):

$$\eta_{PL,T} = \frac{n_T(G_{CS}=0) \cdot k_{f,T}}{G_{LE}} \quad (16)$$

And vice versa for the triplet decay under electro-injection and Equation (5). Under steady-state illumination at open circuit, PM6:Y6 (and other systems with the same offset) loses about 70% of excitations through the triplet state. In contrast, PM6:o-IDTBR exhibits only about 40% triplet losses, mainly attributed to less efficient LE dissociation at smaller offset. This trend is in agreement with experimental observations; 90% triplet losses for PM6:Y6 [14] and around 10-30% triplet losses for PM6:o-IDTBR [55] have been found (see also Supplementary Note 2.9). Triplet-triplet or triplet-charge annihilation mechanisms are not considered in this model and can influence the exact value, especially with high triplet populations [56].

Figure 5b shows a correlation between the increase of electroluminescence (blue) and the decrease of triplet population under electro-injection (purple) for $\Delta E_{LE-} < 0$. However, due to the 3:1 split of free charge recombination, even at very low offsets where the singlet population reaches its maximum, the triplet population is still very significant. In fact, a maximum $\eta_{EL}$ of 25% (which is the "singlet loss") is reached even though the triplet losses make up 75%. Additionally, despite the triplet population quickly dropping at large offset due to less efficient triplet formation and increased triplet re-dissociation, non-radiative losses do not simply cease to play a role. Rather, the increased CT state population leads to substantial non-radiative losses through the CT instead, cf. Ref. [57]. This is visualized in Figure 5c, showing the dominant recombination channel changing with the offset due to the different energetics and subsequent kinetics of LE, CT and T states. Note, that while the non-radiative losses change their origin with offset, dominated by either the triplet or the CT1 state (under illumination also the LE), the total recombination remains constant, as detailed in Equation (S5.8). The presence (absence) of the triplet state therefore does not immediately correlate with high (low) NR losses or low (high) device efficiency. Most modern systems, such as the high-efficiency PM6:Y6, are moderate-offset systems, where the triplets are the primary loss channel.

### 4.4. Free Charge Population

Using a realistic CT formation rate coefficient of $k_{r,CT} = 2.5 \cdot 10^{-1}$ cm$^3$s$^{-1}$ (from Supplementary Note 2.8), the free charge carrier population under illumination can be determined, as visualized in Figure 5a (green). Naturally, it increases with offset, where singlets and triplets dissociate readily and are formed less efficiently. For two example systems we calculate a free charge population for PM6:Y6 of $2.51 \cdot 10^{22} m^{-3}$ and $2.82 \cdot 10^{21} m^{-3}$ for PM6:oIDTBR. These show excellent agreement with measured values of $2 \cdot 10^{22} m^{-3}$ [4] and $2 \cdot 10^{21} m^{-3}$ [55], respectively. The triplet population reaches similar values to the free charge population, in accordance with Ref. [18].

## 4.5 Predicting Device Performance

We now turn to calculating the device efficiency from this model, relating both the open circuit voltage and short circuit current density to our metrics via:

$$qV_{oc}(\Delta E; E_g) = qV_{oc}^r(E_g) + k_B T \cdot \ln\{PLQY_{PS:A} \cdot \eta_{EL}(\Delta E)\} \quad (17a)$$
$$J_{sc}(\Delta E; E_g) = J_{ph}(E_g) \cdot \eta_{CG}(\Delta E) \quad (17b)$$

where $V_{oc}^r$ represents the radiative voltage limit of $V_{oc}$ and the non-radiative voltage loss is accounted for by the electroluminescence through $\eta_{EL}$, using Rau's reciprocity relation, e.g., Ref. [19]. Setting $PLQY_{PS:A} \approx 2\%$ leads to an $\Delta E$-independent downshift of 100mV. The maximum photocurrent density $J_{ph}$ is converted to the actual $J_{SC}$ by the free charge generation efficiency $\eta_{CG}$.

Another major loss in OSCs is the fill factor, which is not addressed directly in this model. There are two main aspects limiting organic FFs: the field dependence of free charge generation and the transport resistance [58] [59]. The former is much more pronounced for low $\Delta E_{LE-C}$, where the exciton dissociation becomes inefficient as the CT energy is increased with the external voltage. The latter stems from the competition of charge extraction (limited by low mobility) with bimolecular recombination and can be quantified through the figure of merit α, derived in Ref. [60]:

$$FF = \frac{u_{oc} - \ln(0.79 + 0.66\, u_{oc}^{1.2})}{u_{oc} + 1} \quad (18a)$$

$$u_{oc} = \frac{qV_{oc}}{(1+\alpha)k_B T} \quad (18b)$$

We can use our calculated γ as a parameter describing the non-geminate charge recombination in our model, which shows the same trends expected for the field dependence of IGE: quickly increasing for small $\Delta E_{LE-CT}$. We therefore link the Langevin reduction factor to the figure of merit $\alpha$ to calculate the FF according to Equations (18). We demonstrate in Supplementary Note 2.8 how using typical organic solar cell properties, e.g. $k_2 \approx 10^{-11} cm^3/s$ or $\mu \approx 10^{-3} cm^2/Vs$, yields the relation

$$\alpha^2 = 100 \cdot \gamma \quad (18c)$$

providing a nice description of the fill factor. High FFs are therefore achieved for $\gamma \leq 10^{-2}$ ($\alpha \leq 1$), which is reached for high-offset systems (Fig. 4d). A small reduction in recombination is already seen in well phase-separated systems such as PM6:Y6, and many other blends that use well aggregating NFAs (e.g. due to their fluorination), as a result of the geometric confinement of electrons and holes within their respective phases [42] [61].

### Efficiency Limits: Shockley Queisser and 5-State Model

The theoretical efficiency limits of solar cells are calculated in the Shockley Queisser picture by modelling the Sun as a black body [62], emitting a spectral photon flux density (Fig. 6a, solid blue line) of:

$$\varphi_T^{bb}(E) = \frac{2\pi E^2}{h^3 c^2} \cdot \left[\exp\left\{\frac{E}{k_B T}\right\} - 1\right]^{-1} \quad (19)$$

The photon flux density above the bandgap is then,

$$\Phi_{bb}(T) = \int_{E_g}^{\infty} \varphi_T^{bb}(E) \cdot dE \quad (20)$$

whereby the dark saturation and maximum photocurrent density, assuming absorption of 0% below and 100% above the bandgap, are expressed as:

$$J_{sat} = q \cdot \Phi_{bb}(T_c) \quad (21a)$$
$$J_{ph} = q \cdot \theta \cdot \Phi_{bb}(T_s) \quad (21b)$$

with the temperature of the cell $T_c = 300K$ and of the Sun $T_S \approx 6000K$. The contribution of the Sun is reduced by its angular size as seen from Earth $\theta = (R_{sun}/d_{Sun-Earth})^2$ with the Sun's radius $R_{Sun} = 6.957 \cdot 10^8 m$ and its distance to Earth $d_{Sun-Earth} = 1.496 \cdot 10^{11} m$, resulting in the (theoretical) AM0 radiation with a total incident power of $P_{inc} \approx 1.5\ kWm^{-2}$, through

$$P_{inc} = \theta \cdot \int_0^{\infty} \varphi_{T_s}^{bb}(E) \cdot E\, dE \quad (22)$$

The measured AM1.5G solar spectrum (Fig. 6a, dotted blue line) shows only small differences to the black body prediction – mainly signal reduction from atmospheric absorption bands and Rayleigh scattering at higher eV. The photocurrent is calculated using AM1.5G to accurately describe real-world solar cell parameters.

Applying Shockley's equation for an ideal solar cell, e.g. [63]:

$$J(V) = J_{ph} - J_{sat}\left[\exp\left\{\frac{q}{k_B T}V\right\} - 1\right] \quad (23)$$

the radiative voltage limit is calculated as:

$$qV_{oc}^r = k_B T \cdot \ln\left(\frac{J_{ph}}{J_{sat}} + 1\right) \quad (24)$$

$J_{ph}$ and $V_{oc}^r$ provide upper bounds to the $J_{SC}$ and $V_{OC}$ under 1 sun illumination. Figure S9a explores their behaviour for varying bandgap energies $E_g$ and different absorption bandwidths $\Delta E_{abs}$. Notably, $J_{ph}$ exhibits a stark dependence on $\Delta E_{abs}$ up to about 2eV, after which $J_{ph}$ approaches the theoretical limit.

Figure 6a illustrates two different solar cell efficiencies. First, the Shockley-Queisser ultimate efficiency (green), where the photocurrent is supplied as described in Eq. 21b (under AM1.5G) and the efficiency is determined from the maximum power point of the JV curve, Equation (23) - reaching an efficiency of 33.4% at $E_g = 1.34eV$. And second, the calculated 5-state model efficiency (purple), as described in the following.

Using the $V_{oc}^r$ and $J_{ph}$ values from Figure S9a (with an absorption of 90% above the bandgap), the model device efficiency

$$PCE = V_{oc} \cdot J_{sc} \cdot FF \quad (25)$$

can be calculated for different acceptor bandgaps and driving forces. Figure 6b shows the experimental data range very nicely recreated by this calculation (solid line). In addition, a collection of literature values for PCEs of D:A blends (see Supplementary Tables S2 and S3) is shown to fall almost completely in the calculated PCE range between the fill factors of 85% and 30% (shaded area). These values seem to be the upper and lower limit of organic fill factors through Equations (18) when using typical device parameters for OSCs, see Figure 2 in [60].

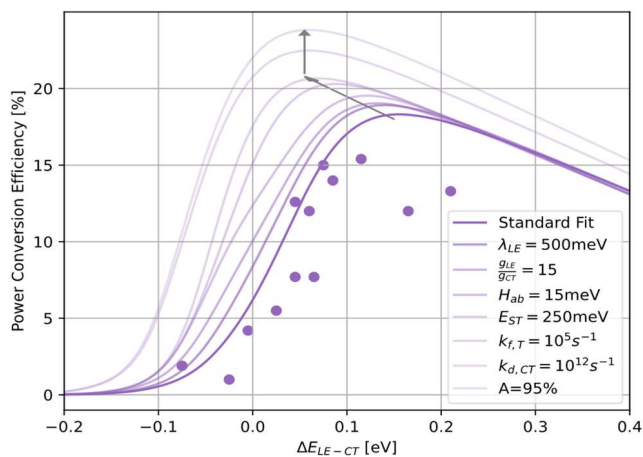

**Figure 7: Increase in PCE from small changes in parameters.** The 'Standard Fit' (same as in Figure 6) uses the values from Table 1, whereas the other lines continuously change one parameter to increase the PCE. Most have a larger effect at small offset.

The PCE curves reveal that the most favourable combination of device parameters can yield a PCE of roughly 20% at an optical acceptor bandgap of $E_{g,A} = 1.4$ eV, with a driving force of $\Delta E_{LE-C} \approx 100-200$ meV. The donor material should have complementary energetics to maximize the combined absorption of the blend, aiming for $\Delta E_{abs} \geq 2$ eV, but is otherwise defined through its HOMO level and consequently $\Delta E_{LE-CT}$. Charge mobility, playing a significant role in extraction efficiency and FF, as well as the proper blend morphology to enable charge generation are, of course, essential but not considered in this model directly.

Figure S8 illustrates how the efficiency of our model varies with different considerations for the CT state. Figure S8a shows the simple 5-state model, disregarding the "deeper CT state" we introduced to resolve the EL bump. Maintaining the same CT decay ($k_{f,CT} = 10^9 s^{-1}$) for both photoexcitation and electro-injection results in an increased $\eta_{EL}$ and consequently an enhanced PCE, particularly for $\Delta E_{LE-CT} > 0.1$ eV, reaching a maximum of 21.4%. Figure S8b looks at the PCE curve if we introduce a CT dissociation which reduces for low-offset as proposed in Ref. [2]. This is discussed further in Supplementary Note 2.12 and shown in Figure S5. It is only noticeable for $\Delta E_{LE-} < 0.1$ eV, where the efficiency is decreases slightly – leaving the maximum PCE unchanged. These two adjustments show how the simple 5-state model can approximate the experimental PCE data, and therefore is a viable model for modern OSCs, while some CT modifications offer valuable improvements.

## 5. Outlook

Our model predicts the PCE of OSCs to be limited to slightly above 20%, which fits nicely with recent record efficiencies for binary blends (Fig. S10). This limit, it should be reiterated, is not an inherent limit to OPVs but stems from the characteristics of modern polymer-donor:NFA binary BHJs. Fullerene devices, for example, are not fitted with these parameters and a different generation of materials could exhibit different behaviour and exceed this limit.

As it stands, there are different approaches to increasing this PCE limit, especially for low-offset systems (see Fig. S6 and Fig. S7). This section explores these approaches, condensed in Figure 7, where the maximum PCE can be seen to be increases quite drastically and, importantly, towards lower energy offset.

### Triplet Excitons

The reduction of $\Delta E_{ST}$, e.g. through increased end-group $\pi$-$\pi$ stacking [41], changes the balance of triplet dissociation and formation, decreasing non-radiative triplet losses and the Langevin reduction factor through higher T and therefore CT3 re-dissociation (Eq. 15). This could e.g. be achieved through the hybridization of T1 and CT3 [14]. Charge up-conversion through triplet-triplet annihilation is additionally affected by the crystallinity of the NFA domains and can become a relevant effect for longer triplet lifetimes $\tau_T \gtrsim 1\mu s$ [56]. Triplet dynamics can improve OPV's FF and reduce voltage losses by tens of meV, especially at low offset. A slight increase of $J_{SC}$ can be expected through lower triplet recombination, but is not seen in our modelled free charge generation pathway.

### Singlet Excitons

A long singlet lifetime is vital for charge generation [9]. The success of the Y6-based devices is partially said to come from its molecular packing, shown to be linked to the electronic coupling as well as excited state reorganization energy [64]. Similarly, singlet exciton losses can be reduced by increased coupling to the donor $H_{ab}$ or reduced reorganization energy $\lambda_{LE}$, e.g. through increased rigidity of the NFA molecule [5]. It may be surprising that these two parameters affect the LE competition (dissociation vs. reformation), which is seemingly independent of both:

$$\frac{k_{r,LE}}{k_{d,LE}} = 10 \cdot \exp\left\{-\frac{\Delta E_{LE-CT}}{k_B T}\right\} \quad (26)$$

However, the *effective* reformation $k_{r,LE}^{eff}$ is decreased by an increased dissociation efficiency $\eta_{d,LE}$, whereby $H_{ab}$ and $\lambda_{LE}$ can indeed have an impact through $k_{d,LE}$. This competition defines the behaviour of the IGE and is therefore detrimental to making low-offset systems work. As evident from Equation (26), a change in the pre-factor can also control the competition through morphology optimization – the pre-factor being defined by phase separation and exciton diffusion lengths in BHJs [31]. We finally note, that recent efficiency records have used increased exciton generation through light-harvesting methods such as a wrinkled AL surface [65] or anti-reflective coating [66].

### CT States

CT dissociation is already assumed to be efficient compared to its decay [8]. Through Langevin reduction as a measure of CT re-dissociation, the FF is influenced by the CT dissociation. Additionally, a disruption of the CT dissociation efficiency will heavily impact the IGE and thereby $J_{SC}$. Counteracting the vanishing CT dissociation at low-offset [2] is important to increase low-offset efficiency but will have little impact on the overall efficiency limit.

### Free Charge Extraction

Finally, the main limit on the FF in record organic solar cells, where field-assisted LE dissociation is not an issue [59], is set by inefficient charge extraction competing with non-geminate recombination. FFs exceeding 80% are very rare. Recently, a certified FF of 80.5% was reported for a ternary blend with a PCE of 20% by means of a dual nanoscale fibrillar morphology through a novel NFA [67].

### Ternary Systems

We finally note, that certified PCEs close to 21% have been recently reported for ternary blends, e.g. Ref. [66]. It is not clear yet what mechanism causes the superior performance of ternary versus binary blends. A recent publication suggested that adding a third component will give a smaller offset which may decrease the exciton dissociation rate to a value where the free charge generation is still efficient while the ELQY benefits from less efficient re-dissociation of LE, thereby increasing the radiative efficiency and reducing the non-radiative voltage loss [19].

## 6. Conclusion

In this work, we developed an analytical framework to comprehensively describe free charge generation and recombination as functions of the D:A energy offset, by introducing a 5-state model that specifically considers the decay and re-dissociation of local triplet states which are populated through electron transfer from the triplet CT state.

We highlight the merits and limitations of models based solely on the singlet branch, contrasting them with those incorporating triplets as well. In this light, we demonstrate the viability of our comprehensive model to accurately reproduce key metrics commonly used to characterize OSCs. Only with the inclusion of triplets can we reconcile both the generation and recombination pathways, offering a unified framework that fits not only IGE and PLQY, but also ELQY and the Langevin reduction factor consistently. These metrics are heavily influenced by the energetics and kinetics of the charge species, including the triplet excitons. Consequently, our model can estimate important parameter values, such as the reorganization energy for electron transport between CT1 states and localized singlets ($\lambda_{LE} \approx 550 meV$) or the singlet-triplet energy offset $\Delta E_{ST} \approx 300 meV$ as well as confirm certain concepts such as the Langevin reduction being a measure of CT (re)dissocation or a broadened DOS playing a differentiating role in generation and recombination. Additionally, we can extract vital information from the population dynamics, such as the magnitude of triplet losses or the free charge carrier density across different systems. Importantly, high triplet losses were found for systems such as PM6:Y6 – a well performing system – and a reduction of triplet losses are not necessarily correlated with an increased ELQY due to the presence of non-radiative CT decay, especially at high offset.

Furthermore, these metrics enable us to recreate device parameters - namely $V_{oc}$, $J_{sc}$, FF and the PCE - supported by SQ calculations and validated with extensive data from the current literature. Our analysis of the device performance has revealed key energetic requirements for both active layer materials to achieve an optimum efficiency of 20.3% for binary blends. These include an acceptor bandgap of 1.4eV, a LE-CT offset of 150meV (roughly a HOMO level difference of 350meV) and complementary absorption of the donor to ensure high photocurrent. Our finding explains the PCE plateau for binary blends observed in recent years, where OSC efficiency has not increased much above 20% if no measures to increase light harvesting are employed.

We also demonstrated different means to achieve higher efficiencies, especially for low-offset systems, such as reducing the singlet-triplet offset, reorganization energy and domain size or increasing the D-A coupling and absorption strengths. Theoretically, an optimized NFA design could incorporate all these factors, leading to a novel class of materials for organic photovoltaics with a higher efficiency limit.

**Acknowledgement**

We acknowledge financial support by the Leibnitz Association and the German Research Foundation (DFG) projects *Fabulous* (project number 450968074) and *Extraordinaire* (project number 460766640).

**Author Contributions**

Safa Shoaee and Dieter Neher conceptualized the study and designed the methodology. Manasi Pranav provided the experimental data, found in the Supplementary Information [68]. Jonathan Langentepe-Kong performed the analysis and wrote the majority of the manuscript.

# SUPPORTING INFORMATION

## Triplet Excitons Reconcile Charge Generation and Recombination in Low-Offset Organic Solar Cells: Efficiency Limits from a 5-State Model

Jonathan L. Langentepe-Kong, Manasi Pranav, Safa Shoaee, Dieter Neher

## 1. Experimental Details and Data

### 1.1 Device Fabrication

Device structure usually follows ITO/PEDOT:PSS/AL/PDINN/Ag.

Glass substrates with pre-patterned ITO (Lumtec) were cleaned in an ultrasonic bath with Hellmanex III, deionized water, acetone and isopropanol for 10min each. Afterwards, they were treated by microwave oxygen plasma at 200W for 4min. An aqueous solution of PEDOT:PSS was filtered through 0.45um PTFE filter and spin coated onto the substrate at 5000rpm under ambient condition, then annealed at 150C for 15min. Blend solutions were prepared with a total concentration of 14mg/mL in Chloroform using D:A weight ratio of 1:1.2. They were stirred for 3h inside a nitrogen-filled glovebox, spin-coated at 3000rpm onto the substrate and annealed at 100C for 10min to result in 100nm active layer thickness. Afterwards, a 1mg/mL solution of PDINN in methanol was spin coated at 2000rpm as the ETL. Lastly, 100nm of silver as the top electrode was evaporated under a $10^{-6} - 10^{-7}$ mbar vacuum. This resulted in a 6mm² BHJ device. For films, the AL was spin-coated onto glass substrates instead, without TLs or electrodes. The exact numbers for concentration, spin speed and annealing may vary slightly for different material combination to result in their highest PCE.

### 1.2 Current Density – Voltage Characteristics (JV)

JV curves were measured using a Keithley2400 in a 2-wire configuration under simulated AM1.5G irradiation at 100mW/m², provided by a filtered Oriel Sol2A Class AA Xenon lamp. The sun simulator was calibrated with a KG5 filtered silicon solar cell (certified: Fraunhofer ISE) and its intensity monitored for each measurement using a Si photodiode. We determine the efficiency:

$$PCE = \frac{J_{SC} V_{OC} FF}{P_{inc}}$$

### 1.3 External Quantum Efficiency (EQE-PV)

EQE illumination was supplied with a 300W Halogen lamp (Phillips), chopped at 80Hz (Thorlabs MC2000), guided through a monochromator (Tornerstone) and coupled into an optical quartz fibre before hitting the sample. The setup is calibrated with Newport Photodiode (818-UV). A lock-in amplifier (SR830) measures the device's response to get:

$$EQE = \frac{\#extracted\ charges}{\#incident\ photons} = \frac{J}{q} \frac{h\nu}{P}$$

The maximum number of extracted charges is limited by the absorbed photons, such that $EQE = absorption * IGE$. We can therefore obtain the IGE as the experimental EQE normalized to ca. 80%, depending on the absorption of each material combination.

### 1.4 Photoluminescence Quantum Yield (PLQY)

PL measurements were performed using a 520nm CW laser diode (Insaneware), guided through an optical fibre into an integrating sphere (Hamamatsu Photonics KK A10094) holding the film. Another optical fibre channelled the output to a spectrograph (Andor Solis SR393i-B). The light intensity is measured with a Si photodiode to provide 1 sun illumination.

$$PLQY = \frac{\#PL\ Emission}{\#Absorption}$$

The PL emission is measured with a filter to reduce the signal from the excitation light. The absorption can be inferred from the intensity difference at excitation wavelength between measurements with and without sample (measurements done without filter). Background corrections are done through dark measurements.

Films were encapsulated with thin glass films beforehand.

## 1.5 Electroluminescence Quantum Yield (ELQY) and temperature dependent ELQY (T-EL)

A Si photodetector (Newport) with ~2cm² area, connected to a picoampere meter (Keithley 485), measured the response of the device to an injection current (equal to the short circuit current but at positive bias) supplied by a Keithley2400. The total photon flux was evaluated considering the device's emission spectrum, the detector's quantum efficiency and the distance-dependent signal drop, summed up a correction factor C.

$$ELQY = \frac{EL\ current\ (Si\ photodetector)}{Injection\ current} \cdot C$$

Temperature dependent measurements were conducted with the device placed inside a nitrogen cryostat, controlled via a Lakeshore (model 335) temperature controller. T- EL probes the thermally activated back-transfer process from CT states to LE states and the data can be used to fit the energy offset $\Delta E_{LE-CT}$ [69].

## 1.6 Time-Delayed Collection Field (TDCF) and Bias-Assisted Charge Extraction (BACE).

A 1mm² device was excited with a 500Hz, ≈5ns pulsed Nd:YAG laser (NT242), homogeneously scattered through an 85m silica fibre (LEONI). A square voltage transient waveform is applied to the device (Aglient 81150A), triggered by a photodiode (EOT, ET-20230TTL), so that the device is held at different pre-biases while photoexcited, whereupon a high reverse bias collects the charges. The current is measured using a 50Ω resistor (in series) and recorded with an oscilloscope (Aglient DSO9104H). Non-geminate recombination is kept to a minimum by using a low fluence of 0.05Jcm$^{-2}$ and very quick bias ramp up (≈ns).

In BACE, the light is supplied by a 520nm diode laser at 500Hz, 50% duty cycle. Under illumination at different light intensities the device is kept at $V_{OC}$ (steady state) before extracting all charges with a high collection bias (in the dark). A homogeneous generation profile is maintained through an AL thickness of d >150nm.

$$k_2 n^2 = R = G = \frac{J_{sc}}{qd}$$

## 1.7 Space-Charge Limited Current (SCLC) and Resistance Dependent Photovoltage (RPV)

Electron and hole-only devices are fabricated in ITO/ZnO/AL/PDINN/Ag and ITO/PEDOT:PSS/AL/MoO3/Ag configuration, respectively. The ZnO nanoparticle dispersion in isopropanol (Avantama N-10), filtered through a 0.45mm filter, was spin coated onto ITO at 5000 rpm (in air) and annealed at 120C for 20 min. MoO3 is evaporated under $10^{-6} - 10^{-7}$ mbar at 0.05-0.1 A/s to attain an 8nm layer. AL thickness is increased above 150nm. The measurement is similar to the JV characteristics under dark conditions (using a Keithley 2400) and evaluated by the Murgatroyd-Gill equation, with field enhancement factor $\gamma_F$:

$$J_{SCLC} = \frac{9}{8}\varepsilon\mu_0 \frac{V^2}{d^3} \exp\left\{0.891\gamma_F \sqrt{\frac{V}{d}}\right\}$$

Alternatively, the mobility $\mu$ can be measured via RPV, using the same Nd:YAG laser and oscilloscope as in TDCF. Photovoltage transients were recorded with a load resistance of 1MΩ. Low laser fluence is used to avoid internal field screening or charge build up.

Mobility and recombination measurements finally lead to the Langevin reduction via:

$$\gamma = \frac{k_2}{k_L} \text{ with } k_L = \frac{q(\mu_n + \mu_p)}{\varepsilon\varepsilon_0}$$

## 1.8 Data from our group used to fit device metrics

**Table S1: BHJ device parameters:** singlet exciton to singlet CT energy offset $\Delta E_{LE-CT}$ [7]; IGE based on $EQE_{PV}$ measurements; the PLQY (of neat acceptor and D:A blend) and ELQY (of the blend) from PLQY and ELQY measurement; the PLQY and ELQY of the blend normalized to the acceptor PLQY; $k_2$ from BACE (and confirmed by TCDF); charge carrier mobilities measured via [5]SCLC or [6]RPV; $\gamma$ based on $k_2$ and $\mu$; $PCE$ from JV measurements. "L" refers to low molecular weight (3.5kDa) PM6 (usually >20kDA).

| Blend [D:A] | $\Delta E_{LE-CT}$ [meV] | IGE [%] | $PLQY_{PS:A}$ | $PLQY_{D:A}$ | $ELQY_{D:A}$ | $\frac{PLQY_{D:A}}{PLQY_{PS:A}}$ | $\frac{ELQY_{D:A}}{PLQY_{PS:A}}$ | $k_2$ [$m^3/s$] | $\mu_{fast}$ [$m^2/Vs$] | $\mu_{slow}$ [$m^2/Vs$] | $\gamma$ | PCE [%] |
|---|---|---|---|---|---|---|---|---|---|---|---|---|
| PM6:Y6 | 115[1,2] | 96 | 0.0103 | 5.2E-4 | 4E-5 | 0.0505 | 3.88E-3 | 1.4E-17 | 5.5E-8[5] | 5.5E-8[5] | 0.0247 | 15.4 |
| PM6:Y5 |  | 60 | 0.0245 | 0.0107 | 2.2E-3 | 0.4367 | 0.0898 | 1E-16 | 1E-7[5] | 1.85E-8[5] | 0.1634 | 7.7 |
| L:Y6 | 25[2] | 65 | 0.0103 | 0.0042 | 3.5E-4 | 0.4078 | 0.0340 | 1.5E-16 | 1E-7[5] | 9.9E-10[5] | 0.2877 | 5.5 |
| L:Y5 | -25[2] | 18 | 0.0245 | 0.0203 | 1.3E-3 | 0.8286 | 0.0531 | / | 1.7E-7[5] | 4E-9[5] | / | 1 |
| PM6:TPT10 | 76[1] | 87 | 0.027 | 0.0016 | 7.4E-4 | 0.0593 | 0.0274 | 5E-17 | / | / | / | 15 |
| PM6:o-IDTBR | -5[3] | 24 | 0.06 | 0.042 | 2E-3 | 0.7 | 0.0333 | 3E-16 | 5E-8[6] | 3.4E-9[6] | 1.0881 | 4.2 |
| PTB7-Th:BTPV-4F-eC9 | 87[1] | 83 | 5.38E-3 | 1.2E-4 | 2.4E-5 | 0.0223 | 4.46E-3 | 2E-17 | 2E-7[5] | 3E-8[5] | 0.0168 | 14 |
| PM6:N4 | 163[1] | 97 | 3.45E-3 | / | 1.2E-6 | / | 3.48E-4 | 2E-18 | 1.6E-8[6] | 1E-9[6] | 0.0228 | 12 |
| PM6:BTPV-4F-eC9 | 47[1] | 43 | 5.38E-3 | 0.0028 | 4.6E-5 | 0.5205 | 8.65E-3 | 6E-17 | / | / | / | 7.7 |
| PTQ10:Y6 | 45[3] |  | 0.0103 | 0.0011 | 9.9E-5 | 0.1068 | 9.61E-3 | / | / | / | / | 12.6 |
| PTQ10:Y5 | -75[3] | 13 | 0.0245 | 0.0201 | 3.8E-3 | 0.8204 | 0.1551 | / | / | / | / | 1.9 |
| PM6:L0 | 60[1] | 85 | 0.0138 | 0.0022 | 1.8E-4 | 0.1594 | 0.0130 | 1E-16 | 2,4E-7[5] | 1.6E-8[5] | 0.0757 | 12 |
| PM6:L4 | 210[4] | 98 | 0.0195 | 3.1E-4 | 1.5E-6 | 0.0160 | 7.59E-5 | 1.4E-17 | / | / | / | 13.3 |

[1]T-EL was used to determine $\Delta E_{LE-C}$.
[2]In situ spectroelectrochemistry (SEC), i.e. coupling of cyclic voltammetry & in situ UV-vis-NIR spectroscopy, was used to determine $\Delta_{HOMO}$. $\beta = 235meV$ was found from PM6:Y6 for the conversion $\Delta E_{LE-CT} = \Delta_{HOMO} - \beta$ [7].
[3]Based on reported HOMO level differences (between PM6 & PTQ10 and o-IDTBR & Y6) and $\beta$, collected in [7].
[4]Based on the energetic comparison between L0 and L4, giving $\Delta_{HOMO} \approx 150eV$ in [70].

## 1.9 Data from the literature used to support the PCE prediction

**Table S2: BHJ device parameters from the literature**, collected in [71]. $\Delta_{HOMO}$ was calculated based on the energy levels reported alongside the PCE or literature values if not given in the reference. Triangles in Figure 6b.

| Blend (BHJ) | $\Delta_{HOMO}$ [meV] | PCE [%] |
|---|---|---|
| PTB7-Th:F0IC | 120 | 12.0 |
| PM6:L8BO | 230 | 18.6 |
| PM6:BTP-eC9 | 190 | 17.8 |
| PtB7-Th:ITIC | 240 | 6.8 |
| PDBT-T1:ITIC-Th | 300 | 9.6 |
| PM6:SeTIC-4Cl | 200 | 13.3 |
| PM6:SeTIC | 100 | 7.5 |
| PM6:Y6 | 200 | 17.1 |
| PM6:BTP-4Cl | 230 | 16.5 |
| PBDB-7:ITIC | 180 | 11.2 |
| D18:Y6 | 140 | 18.2 |
| D18:Y6Se | 190 | 17.7 |

| System | | |
|---|---|---|
| PBDB-T-SF:IT-4F | 260 | 13.1 |
| J91:m-ITIC | 100 | 11.63 |
| D18-Cl:N3 | 90 | 18.13 |
| PM6:IT-4F | 210 | 13.2 |
| PM7:IT-4F | 140 | 14.4 |
| PBDB-TF:IT-4F | 210 | 13.8 |
| T1:IT-4F | 180 | 15.1 |
| T2:IT-4F | 150 | 14.5 |
| T3:IT-4F | 110 | 13.9 |
| PTO2:IT-4F | 60 | 14.5 |
| D16:Y6 | 170 | 16.72 |
| D18:N3 | 140 | 18.56 |
| J61:ITIC | 160 | 9.53 |
| J60:ITIC | 160 | 8.97 |
| PBDB-T:Y1 | 60 | 13.42 |

**Table S3: Record efficiencies of the last 3 years** from the literature ($^c$ certified), also including ternary cells (no offset for ternaries), plotted in Figure S10 and as pluses in Figure 6b. No.3 (bold) is the marked outlier, exceeding PCE expectations through a light-enhancing surface roughness. No.16 not visible in Figure 6b due to a too negative offset.

| No. | System | $\Delta_{HOMO}$ [meV] | PCE [%] | Reference |
|---|---|---|---|---|
| 1 | D18:N3-BO:F-BTA3 | | 19.82 $^c$ | [72] |
| 2 | SA1:PFDTQ:PY-IT | | 19.17 $^c$ | [73] |
| 3 | **D18-Cl/BTP-4F-P2EH** (as a BHJ) | 70 | **20.10** $^c$ (18.4) | [65] |
| 4 | D18:L8-ThCl/L8-BO:L8-ThCl | | 20.00 $^c$ | [67] |
| 5 | PM6:D18-Cl:NA3 | | 19.39 $^c$ | [74] |
| 6 | PM6:BTP-eC9:SMA | | 19.66 $^c$ | [75] |
| 7 | D18:Z8:L8-BO | | 19.8 $^c$ | [76] |
| 8 | D18:LJ1:BTP-eC9-4F | | 19.43 | [77] |
| 9 | D18:LJ1:L8-BO | | 19.78 | |
| 10 | D18-Fu:L8-BO | 200 | 19.11 | [78] |
| 11 | PM6:BTP-BO-4FO (CN) | 130 | 18.62 | [79] |
| 12 | PM6:L8-BO:BTP-eC9 | | 19.71 | [80] |
| 13 | PBDB-TF:L8-BO:BTP-eC9 | | 19.79 $^c$ | [81] |
| 14 | PM6:BTP-eC9 with BrBACz HTL | 190 | 19.7 | [82] |
| 15 | D18:PM6:L8-BO | | 19.9 | [83] |
| 16 | D18:AQx-2F | -370 | 19.7 | [84] |
| 17 | HW-D18:L8BO | 115 | 19.65 | [85] |
| 18 | PM6:BTP-C9:o-BTP-eC9 | | 19.5 $^c$ | [86] |
| 19 | D18:DT-C8Cl | 270 | 19.4 | [87] |
| 20 | PM6:L8-BO:YR-SeNF | | 18.8 | [88] |
| 21 | PM1:A-OSeF | 220 | 18.5 | [89] |
| 22 | D18:D18-Cl:CH8F | | 19.28 | [90] |
| 23 | PM6:L8-BO | 230 | 19.02 | [91] |
| 24 | PM6:BTP-eC9 | 190 | 18.93 $^c$ | [92] |
| 25 | D18:BTP-Cy-4F:BTP-eC9 | | 19.36 | [93] |
| 26 | PM6:PM7-Si:BTP-eC9 | | 18.9 | [94] |
| 27 | PM6:D18:L8-BO | | 19.2 $^c$ | [95] |
| 28 | PM6:L8-BO | 230 | 19.1 | [96] |
| 29 | D18:L8-BO | 170 | 19.3 | |
| 30 | PM6:L8-BO:B6Cl | | 19.8 | |
| 31 | D18:L8-BO:B6Cl | | 20.2 | |
| 32 | PM6:L8-BO-C4 | 210 | 19.78 | [97] |
| 33 | PM6:L8-BO-C4:L8-BO-C4-Br | | 20.2 $^c$ | |

## 2. Analytical Description of the State Models

**2.1 The 2-state model** only relies on the LE and CT state, with the additional CT dissociation $k_{d,CT}$. The system of equations in steady state is given as:

$$\begin{bmatrix} -(k_{d,LE} + k_{f,LE}) & k_{r,LE} \\ k_{d,LE} & -(k_{r,LE} + k_{d,CT} + k_{f,CT}) \end{bmatrix} \begin{bmatrix} n_{LE} \\ n_{CT} \end{bmatrix} = -\begin{bmatrix} G_{LE} \\ G_{CT} \end{bmatrix} \quad (S1.1)$$

with the solution:

$$n_{LE} = \frac{G_{LE}(k_{r,LE} + k_{d,CT} + k_{f,CT}) + G_{CT} k_{r,LE}}{k_{d,LE}(k_{d,CT} + k_{f,CT}) + k_{f,LE}(k_{r,LE} + k_{d,CT} + k_{f,CT})} \quad (S1.2)$$

$$n_{CT} = \frac{G_{LE} k_{d,LE} + G_{CT}(k_{d,LE} + k_{f,LE})}{k_{d,LE}(k_{d,CT} + k_{f,CT}) + k_{f,LE}(k_{r,LE} + k_{d,CT} + k_{f,CT})} \quad (S1.3)$$

The direct generation into the CT is usually negligible under 1 sun illumination [11], $G_{CT} = 0$.

**2.2 The metrics** can be inferred from these state populations. IGE comes from the dissociation of CT states under photo illumination:

$$\eta_{CG} = \frac{n_{CT}(G_{CT} = 0) \cdot k_{d,CT}}{G_{LE}} = \frac{k_{d,LE} \cdot k_{d,CT}}{k_{d,LE}(k_{d,CT} + k_{f,CT}) + k_{f,LE}(k_{r,LE} + k_{d,CT} + k_{f,CT})} \quad (S2.1)$$

which yields the same result as Equation 3 in the main text. As a metric relying solely on the singlet branch and LE-CT interplay, ignoring back transfer from CS, it holds for the more complex models as well, as seen in Eq. S6.5.

We have defined the PL decay efficiency $\eta_{PL}$ as the fraction of excitations that decay through the LE state. The luminescence is only the radiative part of this, such that

$$\text{PLQY}_{D:A} = \eta_{PL} \cdot \frac{k_f^r}{k_f} \quad (S2.2)$$

This radiative fraction can now be identified as:

$$\frac{k_f^r}{k_f} = \frac{k_f^r}{k_f^r + k_f^{nr}} = \text{PLQY}_A \quad (S2.3)$$

where the total decay rate coefficient is split into the radiative and non-radiative parts, $k_f^r$ and $k_f^{nr}$, respectively. Strictly speaking, an additional term $C$ should be introduced in both equations to represent the detection efficiency of each photon. This includes the probability of the photon escaping the device to reach the detector without being reflected, reabsorbed or scattered; and the detector's quantum efficiency. However, since both measurement are performed under identical conditions – (notably, the integrating sphere) – this detection efficiency $C$ is the same in both cases and cancels out when the PLQY's are divided in Equation (6).

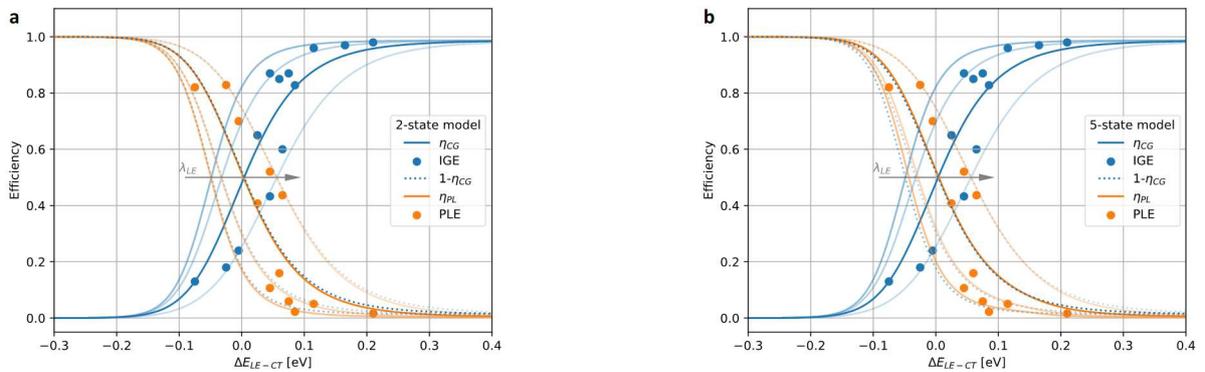

Figure S1: Similar to Fig. 2a but 2-state and 5-state models viewed separately. Through comparison of $\eta_{PL}$ with $1 - \eta_{CG}$ (dotted) the competition between singlet decay and free charge generation can be seen very well.

**2.3 The 3-state model** adds the CS state with reformation to the CT.

$$\begin{bmatrix} -(k_{d,LE} + k_{f,LE}) & k_{r,LE} & 0 \\ k_{d,LE} & -(k_{r,LE} + k_{d,CT} + k_{f,CT}) & k_{r,CT} \\ 0 & k_{d,CT} & -k_{r,CT} \end{bmatrix} \begin{bmatrix} n_{LE} \\ n_{CT} \\ n_{CS} \end{bmatrix} = -\begin{bmatrix} G_{LE} \\ 0 \\ G_{CS} \end{bmatrix} \quad (S3.1)$$

The generation pathways do not include direct CT excitation. The solution is:

$$n_{LE} = \frac{G_{LE}(k_{r,LE} + k_{f,CT}) + G_{CS} k_{r,LE}}{k_{d,LE} k_{f,CT} + k_{f,LE}(k_{r,LE} + k_{f,CT})} \quad (S3.2)$$

$$n_{CT} = \frac{G_{LE} k_{d,LE} + G_{CS}(k_{d,LE} + k_{f,LE})}{k_{d,LE} k_{f,CT} + k_{f,LE}(k_{r,LE} + k_{f,CT})} \quad (S3.3)$$

$$n_{CS} = \frac{G_{LE} k_{d,LE} k_{d,CT} + G_{CS}[k_{d,LE}(k_{d,CT} + k_{f,CT}) + k_{f,LE}(k_{r,LE} + k_{d,CT} + k_{f,CT})]}{k_{r,CT}[k_{d,LE} k_{f,CT} + k_{f,LE}(k_{r,LE} + k_{f,CT})]} \quad (S3.4)$$

It is interesting to see the changes from the 2-state model to the 3-state model in $n_{LE}$ and $n_{CT}$. Mathematiccally, we have a simple transformation of the CT dissociation $k_{d,CT} \to k_{d,CT}^{eff}$ where the dissociation is effectively reduced by the back transfer, i.e. reformation efficiency $k_{d,CT}^{eff} = [1 - \eta_{r,CT}] \cdot k_{d,CT}$. Since no competing rates on CS exist, however, $\eta_{r,CT} = k_{r,CT}/k_{r,CT} = 1$ whereby the CT dissociation disappears (see also Supplementary Note 2.9).

However, the CS population is affected by both $k_{d,CT}$ and $k_{r,CT}$, their competition defining $n_{CS}$. Equation (S2.1) still holds, since we do not consider back transfer from CS in the charge generation pathway.

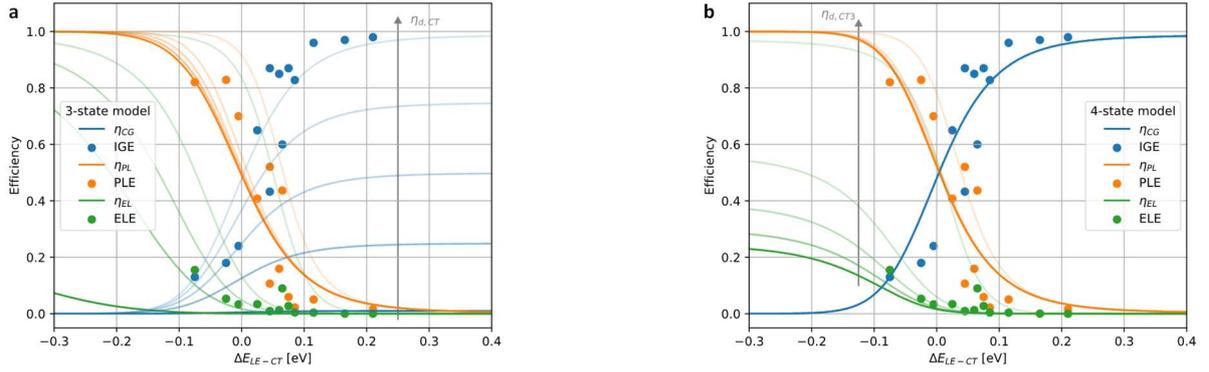

Figure S2: a) Complement to Fig. 2b. IGE, PLE and ELE calculated in the 3-state model for CT1 dissociation efficiencies $\eta_{d,CT} = 0.01, 0.25, 0.5, 0.75, 0.99$. No single value is able to fit all metrics at once. $\eta_{d,CT} \approx 1$ is needed for IGE while PLE and ELE require much smaller values. b) Complement to Fig. 3a. IGE, PLE and ELE in the 4-state model and varying CT3 dissociation efficiencies $\eta_{d,CT} = 0.01, 0.25, 0.5, 0.75, 0.99$. Note: $\eta_{d,CT3} = 1$ is basically the same as $\eta_{d,CT} = 1$, as described in the main text.

**2.4 The 4-state model** adds the CT3 state with CT3 dissociation and decay competition. It is populated with 75% of charge encounters from CS.

$$\begin{bmatrix} -(k_{d,LE} + k_{f,LE}) & k_{r,LE} & 0 & 0 \\ k_{d,LE} & -(k_{r,LE} + k_{d,CT} + k_{f,CT}) & k_{r,CT} & 0 \\ 0 & k_{d,CT} & -4k_{r,CT} & k_{d,CT3} \\ 0 & 0 & 3k_{r,CT} & -(k_{d,CT3} + k_{f,CT3}) \end{bmatrix} \begin{bmatrix} n_{LE} \\ n_{CT1} \\ n_{CS} \\ n_{CT3} \end{bmatrix} = -\begin{bmatrix} G_{LE} \\ 0 \\ G_{CS} \\ 0 \end{bmatrix} \quad (S4.1)$$

Again, no direct CT generation. Solution:

$$N \cdot n_{LE} = G_{LE}\{3k_{d,CT} k_{f,CT3} + (k_{r,LE} + k_{f,CT})(k_{d,CT3} + 4k_{f,CT3})\} + G_{CS} k_{r,LE}(k_{d,CT3} + k_{f,CT3}) \quad (S4.2)$$

$$N \cdot n_{CT1} = G_{LE} k_{d,LE}[k_{d,CT3} + 4k_{f,CT3}] + G_{CS}(k_{d,LE} + k_{f,LE})(k_{d,CT3} + k_{f,CT3}) \quad (S4.3)$$

$$Nk_{r,CT} \cdot n_{CS} = (k_{d,CT3} + k_{f,CT3})\{G_{LE} k_{d,LE} k_{d,CT} + G_{CS}[k_{d,LE}(k_{d,CT} + k_{f,CT}) + k_{f,LE}(k_{r,LE} + k_{d,CT} + k_{f,CT})]\} \quad (S4.4)$$

$$N \cdot n_{CT3} = 3\{G_{LE} k_{d,LE} k_{d,CT} + G_{CS}[k_{d,LE}(k_{d,CT} + k_{f,CT}) + k_{f,LE}(k_{r,LE} + k_{d,CT} + k_{f,CT})]\} \quad (S4.5)$$

$$N = k_{d,CT3}[k_{d,LE}k_{f,CT} + k_{f,LE}(k_{r,LE} + k_{f,CT})] + k_{f,CT3}[k_{d,LE}(3k_{d,CT} + 4k_{f,CT}) + k_{f,LE}(4k_{r,LE} + 3k_{d,CT} + 4k_{f,CT})] \quad (S4.6)$$

Since we now introduced a second rate from CS, the formulas reintroduce the CT1 dissociation due to a reformation efficiency $\eta_{r,CT} < 1$: the equations when $k_{d,CT3} = 0$ are equivalent to those of the 2-state model (in $n_{LE}$ and $n_{CT1}$) with the modification $k_{d,CT} \to k_{d,CT}^{eff} = 75\% \, k_{d,CT}$ since all charges formed into CT3 are lost, $\eta_{r,CT} = 25\%$. On the other hand, the case $k_{f,CT3} = 0$ is equivalent to the 3-state model (in $n_{LE}$, $n_{CT1}$ and $n_{CS}$), since the charge recombination happens exclusively through the CT1, with all CT3 redissociating, $\eta_{r,CT} = 1$.

**2.5 The 5-state model** is described in detail in the main text. Equation (1) is written as:

$$\begin{bmatrix} -(k_{d,LE} + k_{f,LE}) & k_{r,LE} & 0 & 0 & 0 \\ k_{d,LE} & -(k_{r,LE} + k_{d,CT} + k_{f,CT}) & k_{r,CT} & 0 & 0 \\ 0 & k_{d,CT} & -4k_{r,CT} & k_{d,CT3} & 0 \\ 0 & 0 & 3k_{r,CT} & -(k_{d,CT3} + k_{f,CT3}) & k_{d,T} \\ 0 & 0 & 0 & k_{r,T} & -(k_{d,T} + k_{f,T}) \end{bmatrix} \begin{bmatrix} n_{LE} \\ n_{CT1} \\ n_{CS} \\ n_{CT3} \\ n_T \end{bmatrix} = - \begin{bmatrix} G_{LE} \\ 0 \\ G_{CS} \\ 0 \\ 0 \end{bmatrix} \quad (S5.1)$$

Note, that $k_{r,CT}$ can still take on an arbitrary value due to all CS needing to encounter without the competition with a decay channel. CT1 reformation $k_{r,CT}$ only competes with CT3 reformation $3k_{r,CT}$.

The solution is:

$$N \cdot n_{LE} = G_{LE}[k_{d,CT3}(k_{d,T} + k_{f,T})(k_{r,LE} + k_{f,CT}) + k_{r,T}k_{f,T}(4k_{r,LE} + 3k_{d,CT} + 4k_{f,CT})] + G_{CS}k_{r,LE}[k_{d,CT3}(k_{d,T} + k_{f,T}) + k_{r,T}k_{f,T}] \quad (S5.2)$$

$$N \cdot n_{CT1} = G_{LE}k_{d,LE}[k_{d,CT3}(k_{d,T} + k_{f,T}) + 4k_{r,T}k_{f,T}] + G_{CS}(k_{d,LE} + k_{f,LE})[k_{d,CT3}(k_{d,T} + k_{f,T}) + k_{r,T}k_{f,T}] \quad (S5.3)$$

$$Nk_{r,CT} \cdot n_{CS}^2 = [k_{d,CT3}(k_{d,T} + k_{f,T}) + k_{r,T}k_{f,T}]\{G_{LE}k_{d,LE}k_{d,CT} + G_{CS}[k_{d,LE}(k_{d,CT} + k_{f,CT}) + k_{f,LE}(k_{r,LE} + k_{d,CT} + k_{f,CT})]\} \quad (S5.4)$$

$$N \cdot n_{CT} = 3(k_{d,T} + k_{f,T})\{G_{LE}k_{d,LE}k_{d,CT} + G_{CS}[k_{d,LE}(k_{d,CT} + k_{f,CT}) + k_{f,LE}(k_{r,LE} + k_{d,CT} + k_{f,CT})]\} \quad (S5.5)$$

$$N \cdot n_T = 3k_{r,T}\{G_{LE}k_{d,LE}k_{d,CT} + G_{CS}[k_{d,LE}(k_{d,CT} + k_{f,CT}) + k_{f,LE}(k_{r,LE} + k_{d,CT} + k_{f,CT})]\} \quad (S5.6)$$

$$N = k_{d,CT3}(k_{d,T} + k_{f,T})[k_{d,LE}k_{f,CT} + k_{f,LE}(k_{r,LE} + k_{f,CT})] + k_{r,T}k_{f,T}[k_{d,LE}(3k_{d,CT} + 4k_{f,CT}) + k_{f,LE}(4k_{r,LE} + 3k_{d,CT} + 4k_{f,CT})] \quad (S5.7)$$

The total recombination to $S_0$ balances the generation:

$$R_{tot} = n_{LE}k_{f,LE} + n_{CT1}k_{f,CT} + n_T k_{f,T} = G_{LE} + G_{CS} \quad (S5.8)$$

which holds for each pathway independently, i.e. $R_{tot}(G_{CS} = 0) = G_{LE}$ and vice versa. This can be inferred from Figure 5a & b: notice how $n_{LE}$ as well as $n_{CT}$ cross at $n_T/1000$ exactly at $0.5 \cdot 10^{19} m^{-3}$ (where $n_{LE}^{max} = G/k_{f,LE} = 10^{19} m^{-3}$).

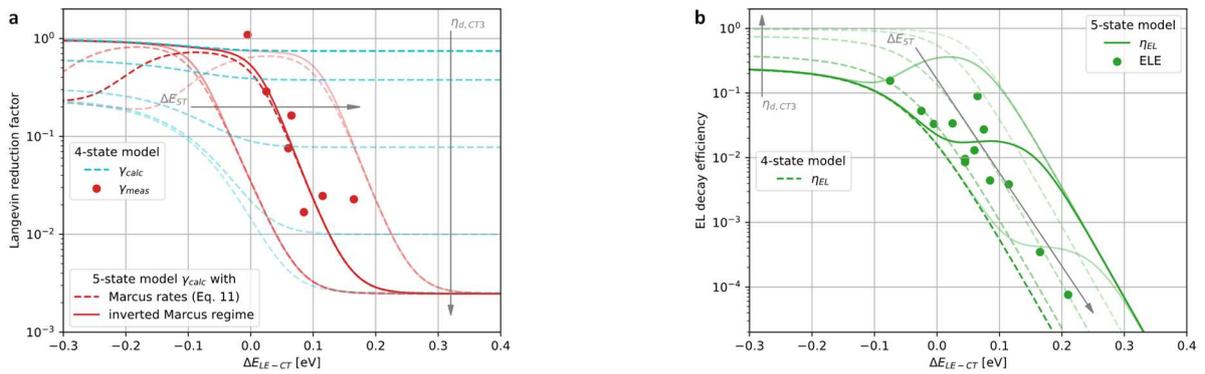

Figure S3: a) The Langevin reduction factor calculated in the 4-state model (blue dotted lines), with varying CT3 dissociation efficiencies $\eta_{d,CT} = 0.01, 0.25, 0.5, 0.75, 0.99$ (vertical arrow). Also in the 5-state model (red lines), for varying singlet-triplet offset $\Delta E_{ST}/meV = 200, 300, 400$ (horizontal arrow). The triplet exciton formation & dissociation is modelled as Marcus rates with (solid) and without (dashed) a constant inverted regime. b) The EL decay efficiency calculated in the 4-state (dotted lines) and 5-state (solid lines) models, with varying $\eta_{d,CT3}$ and $\Delta E_{ST}$ as before (indicated by the arrows). The EL bump is a result of the varying transfer rate coefficients, which lead the model to go through different $\eta_{d,CT}$ values.

## 2.6 CT population rate coefficients

We solve Equations (1a) and (1b) for the CT1 population (in steady-state, $\frac{dn_x}{dt} = 0$):

$$n_{LE} = \frac{G_{LE} + n_{CT1} \cdot k_{r,LE}}{k_{d,LE} + k_{f,LE}} = [G_{LE} + n_{CT1} \cdot k_{r,LE}] \cdot \frac{\eta_{d,LE}}{k_{d,LE}} \tag{S6.1}$$

$$n_{CT1} = \frac{G_{LE} \cdot \eta_{d,LE} + n_{CS}^2 \cdot k_{r,CT}}{(k_{r,LE} + k_{d,CT} + k_{f,CT}) - k_{r,LE}\eta_{d,LE}}$$

$$= \frac{G_{LE} \cdot \eta_{d,LE} + n_{CS}^2 \cdot k_{r,CT}}{(k_{r,LE}^{eff} + k_{d,CT} + k_{f,CT})}$$

$$= [G_{LE} \cdot \eta_{d,LE} + n_{CS}^2 \cdot k_{r,CT}] \cdot \frac{\eta_{d,CT}}{k_{d,CT}} \tag{S6.2}$$

Using $\eta_{d,CT} = \frac{k_{d,CT}}{(k_{r,LE}^{eff} + k_{d,CT} + k_{f,CT})}$ where $k_{r,LE}^{eff} = [1 - \eta_{d,LE}]k_{r,LE}$ with $\eta_{d,LE} = \frac{k_{d,LE}}{k_{d,LE} + k_{f,LE}}$ like in the main text.

Similarly, we solve Equations (1d) and (1e) for the CT3 population (in steady state):

$$n_T = n_{CT3} \cdot \frac{k_{r,T}}{k_{d,T} + k_{f,T}} = [n_{CT3} \cdot k_{r,T}] \cdot \frac{\eta_{d,T}}{k_{d,T}} \tag{S6.3}$$

$$n_{CT3} = n_{CS}^2 \cdot \frac{3k_{r,CT}}{(k_{r,T} + k_{d,CT3}) - k_{r,T}\eta_{d,T}}$$

$$= n_{CS}^2 \cdot \frac{3k_{r,CT}}{(k_{r,T}^{eff} + k_{d,CT3})}$$

$$= [n_{CS}^2 \cdot 3k_{r,CT}] \cdot \frac{\eta_{d,CT3}}{k_{d,CT3}} \tag{S6.4}$$

using $\eta_{d,CT3} = \frac{k_{d,CT3}}{(k_{r,T}^{eff} + k_{d,CT3})}$ where $k_{r,T}^{eff} = [1 - \eta_{d,T}]k_{r,T}$ with $\eta_{d,T} = \frac{k_{d,T}}{k_{d,T} + k_{f,T}}$ to mirror the singlets.

Inserting these Equations (S6.2) and (S6.4) into Equation (1c) we have:

$$\frac{dn_{CS}}{dt} = G_{CS} + [G_{LE} \cdot \eta_{d,LE} + n_{CS}^2 \cdot k_{r,CT}]\eta_{d,CT} - n_{CS}^2 \cdot (1+3)k_{r,CT} + [n_{CS}^2 \cdot 3k_{r,CT}] \cdot \eta_{d,CT3}$$

$$= G_{CS} + G_{LE}\eta_{CG} - n_{CS}^2 \cdot [(1 - \eta_{d,CT}) + 3(1 - \eta_{d,CT3})]k_{r,CT} \tag{S6.5}$$

We can identify the free charge generation pathway through photoexcitation as $G_{LE}\eta_{CG}$, with generation efficiency $\eta_{CG} = \eta_{d,LE} \cdot \eta_{d,CT}$ as described in 2.2, Equation (S2.1) and Equation (3) in the main text.

We get the free charge population, in steady-state:

$$n_{CS}^2 = \frac{G_{CS} + G_{LE}\eta_{CG}}{k_{r,CT}^{eff} + 3k_T} \tag{S6.6}$$

With the CT1 population rate coefficient $k_{r,CT}^{eff} = k_{r,CT}[1 - \eta_{d,CT}]$ from the effective CS encounter to CT1. We similarly introduce the CT3 population rate coefficient $k_T$, holding all triplet terms:

$$k_T = k_{r,CT}[1 - \eta_{d,CT3}]$$

$$= k_{r,CT} - \frac{k_{r,CT} \, k_{d,CT3}(k_{d,T} + k_{f,T})}{k_{r,T}k_{f,T} + k_{d,CT3}(k_{d,T} + k_{f,T})} \tag{S6.7}$$

This rate coefficient - which is the origin of all $k_{r,T}k_{f,T}$ and $k_{d,CT3}(k_{d,T} + k_{f,T})$ terms, as well as of all numerical pre-factors 3 and 4 in the population density solutions (Eq. S5.2 – S5.7) - changes quite drastically, if any one of the triplet rates (through $k_{d,CT3}, k_{d,T}, k_{f,T}, k_{r,T}$) is neglected:

- For $k_{f,T} = 0$, the decay channel through T disappears. In steady state, the triplet formation and dissociation rates compensate each other, whereby the triplet excitons become irrelevant. Then, CT3 formation and dissociation must compensate each other (leading to $\eta_{d,CT3} = 1$), whereby the CT3 become irrelevant. We reduce the system to a 2-state model, similar to what is discussed at the end of Sec. 2.4.
- For $k_{r,T} = 0$, the decay channel through T cannot be reached, which has the same effect.
- For $k_{d,CT3} = 0$, no charges that enter the triplet pathway (through CT3) can repopulate CS, rendering it an effective decay channel. We get $\eta_{d,CT3} = 0$. The system effectively reduces to a 3 state-model with a direct decay pathway $3k_{r,CT} \cdot n_{CS}^2$ from CS (different from Sec 2.3!) – triplets becoming irrelevant.
- Similarly, for $k_{d,T} = 0$, triplet excitons become a dead-end. The system effectively reduces to a 4 state-model with a CT3 decay pathway $k_{r,T}$ (the same as Sec. 2.4 but the decay is not constant!).

This is summarized as follows:

$$k_T \to \begin{cases} 0 & ; k_{f,T} = 0 \text{ or } k_{r,T} = 0, \to \text{2-state model} \\ k_{r,CT} & ; k_{d,CT3} = 0 \to \text{3-state model} \\ k_{r,CT}\left(1 - \dfrac{k_{d,CT3}}{k_{r,T} + k_{d,CT3}}\right) & ; k_{d,T} = 0 \to \text{4-state model} \end{cases} \quad (S6.8)$$

**2.7 The Langevin reduction factor**, by inserting Equations (S5.7) & (S5.4) into Equation (10), becomes

$$\gamma = \frac{k_2}{k_L} = \frac{G_{CS}}{4k_{r,CT} \cdot n_{CS}^2(G_{LE} = 0)}$$
$$= \frac{k_{d,CT3}(k_{d,T} + k_{f,T})[k_{d,LE}k_{f,CT} + k_{f,LE}(k_{r,LE} + k_{f,CT})] + k_{r,T}k_{f,T}[k_{d,LE}(3k_{d,CT} + 4k_{f,CT}) + k_{f,LE}(4k_{r,LE} + 3k_{d,CT} + 4k_{f,CT})]}{4[k_{d,CT3}(k_{d,T} + k_{f,T}) + k_{r,T}k_{f,T}][k_{d,LE}(k_{d,CT} + k_{f,CT}) + k_{f,LE}(k_{r,LE} + k_{d,CT} + k_{f,CT})]} \quad (S7.1)$$

Another description of $\gamma$ can be found from Equation S6.5 by identifying the total recombination $R_{tot} = G_{CS}^{tot} = G_{CS} + G_{LE} \cdot \eta_{CG}$ and the Langevin recombination rate coefficient $k_L = 4k_{r,CT}$. Thereby, the term in the brackets must be (four times) the Langevin reduction factor so that the bimolecular recombination is $R_2 = n_{CS}^2 \cdot \gamma \cdot k_L$. We get $\gamma$, expressed in terms of the CT dissociation efficiencies:

$$\gamma = \left[\frac{1}{4}(1 - \eta_{d,CT}) + \frac{3}{4}(1 - \eta_{d,CT3})\right] \quad (S7.2)$$

It can very easily be seen how the Langevin reduction is a measure of the re-dissociation of CT states.

We also get a quick look at the importance of triplet excitons and the corresponding rate coefficients:
$k_{f,T} = 0$ (or $k_{r,T} = 0$) resulted in $\eta_{d,CT3} = 1$ which would reduce $\gamma$ significantly (upper limit of ¼), whereas $k_{d,CT3} = 0$ yielded $\eta_{d,CT3} = 0$ which would increase $\gamma$ significantly (lower limit of ¾). This is also true for the singlet branch, where higher dissociation of CT – through reduced back-transfer from CT1 ($k_{r,LE}$) as well as reduced decay $k_{f,CT}$ – would ultimately lead to much suppressed recombination and higher charge generation, i.e. higher FF and current [25]. More in the Outlook and Figure S6 and Figure *S7*.

**2.8 Fill Factor considerations** are based on the Figure of Merit α from [60]. It can be given in terms of measurable device parameters:

$$\alpha^2 = \frac{q}{4(k_BT)^2} \frac{k_2 J_G d^3}{\mu_n \mu_p} = \frac{1}{400} \frac{d[nm]^3 * J_{SC}\left[\frac{mA}{cm^2}\right] * k_2\left[\frac{cm^3}{s}\right]}{\mu_n\left[\frac{cm^2}{Vs}\right] \mu_p\left[\frac{cm^2}{Vs}\right]} \quad (S8.1)$$

and links to the FF through Eq. (18):

$$FF = \frac{u_{oc} - \ln(0.79 + 0.66 u_{oc}^{1.2})}{u_{oc}+1} \text{ with } u_{oc} = \frac{qV_{oc}}{(1+\alpha)k_BT} \quad (S8.2)$$

Such that α < 1 is needed to reach a high FF ≥ 80%. Using common device parameters,

$$\alpha^2(d = 100, J_{SC} = 20, k_2 = 10^{-11}, \mu = 7*10^{-4}) = 1 \quad (S8.3)$$

With these values, we can infer the Langevin recombination (using ε = 3.5, $\mu_n = \mu_p = \mu$):

$$k_L = \frac{q}{\varepsilon_0 \varepsilon_r}(\mu_n + \mu_p) \approx 10^{-9} \frac{cm^3}{s} = 100 \cdot k_2 \quad (S8.4)$$

Such that

$$\alpha^2(d = 100, J_{SC} = 20, k_2 = \gamma k_L, \mu = 7*10^{-4}) = 100 \cdot \gamma \quad (S8.5)$$

We can also infer the CT reformation rate from $k_L$ (Eq. 8): $k_{r,CT} = \frac{k_L}{4} = 2.5 \cdot 10^{-16} \frac{cm^3}{s}$, leading to slight overestimation of $n_{CS}$ (Sec. 4.4). A Langevin reduction factor of $\gamma < 0.01$ is therefore needed to reach high FFs, above which the FF reduces with $\alpha = 10\sqrt{\gamma}$. We have $\alpha \propto \mu_{eff}^{-1} = \frac{\mu_n + \mu_p}{\mu_n \cdot \mu_p}$, so that higher mobilities result in a lower pre-factor than 100 and higher FF, whereas lower mobilities result in smaller FF, as expected.

## 2.9 Determination of the fraction of free charge recombination proceeding through the triplet state of PM6:o-iDTBr

In Ref. [55], this fraction has been determined to be ca. 10 %. Here, we take an alternative approach to calculate the triplet density from the reported PIA data and the corresponding contribution of triplet recombination to the total recombination current.

Starting with the reported molar extinction coefficient of solid-state neat o-IDTBR, $\epsilon = 1.2 \times 10^5 M^{-1} cm^{-1}$ [98], we calculate the $S_0$-$S_1$ absorption cross section to be:

$$\sigma[cm^2] = \frac{\epsilon \left[\frac{liter}{mol\,cm}\right]}{N_A \left[\frac{molecules}{mol}\right] \times 0.001 \frac{liter}{cm^3}} = 2 \times 10^{-16}\, cm^2 \qquad (S9.1)$$

Note that $\sigma$ as calculated above is the "decadic" cross section according to $T = 10^{-(\sigma n d)}$ and not $T = e^{-(\sigma n d)}$ as often used.

According to Figure 2e in Ref. [55], the PIA of the $T_1$-$T_n$ triplet absorption in sensitized neat o-IDTBR is ca. 1.1 times larger than of the $S_0$-$S_1$ transition. Therefore, we estimate the triplet PIA cross section to $\sigma_T \approx 2.2 \times 10^{-16}\, cm^2$.

For the illuminated annealed PM6:o-IDTBR blend, PIA yields $\frac{\Delta T}{T} = 4.5 \times 10^{-5}$ for the triplet-associated peak at 950 nm (Figure 2g in Ref. [55]). Then, with

$$\frac{\Delta T}{T} = \sigma_T[cm^2] \times \frac{N_T}{A}\left[\frac{triplet\ excited\ molecules}{cm^2}\right] \times ln10 \qquad (S9.2)$$

the density of molecules in the triplet state per unit area is calculated to $\frac{N_T}{A} = 8.9 \times 10^{10}\, \frac{triplet\ excited\ molecules}{cm^2}$. Finally, the recombination current density through the o-IDTBR triplet state is:

$$J_{R,T} = e \frac{N_T}{A}\left[\frac{triplet\ excited\ molecules}{cm^2}\right] k_{T,d} = 1.4 \frac{mA}{cm^2} \qquad (S9.3)$$

Here, $k_{T,d} = 10^5\, s^{-1}$ is the triplet decay rate. To compare this value with the total recombination current of the blend at $V_{OC}$, we also take into account that charge generation is field dependent in this blend. According to [58], the charge generation efficiency at $V_{OC}$ is ca. 70 % of that at SC, yielding a charge generation current density at $V_{OC}$ of ca. $J_{G,OC}^{1sun} \approx 5.2\, mA/cm^2$.

Therefore, the fraction of charge recombination through triplet is ca. 27 %. While this is somewhat larger than reported in [55], it is still significantly smaller than the CT recombination, in full agreement to the prediction of our model.

**2.10 CT emission as contribution to PL and EL** is neglected in the main text since the CT is less radiative than the LE by three to five orders of magnitude [57]. Figure S4 shows the effect of different ratios of LE to CT radiative fraction. A usual fraction of $10^{-3}$ does not change the PL or EL decay efficiency significantly.

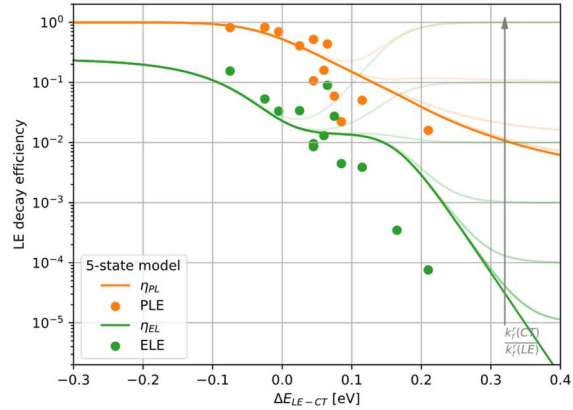

Figure S4: PL and EL decay efficiencies (calculated in the 5-statw model vs. experimental) when considering a radiative CT state, using different radiative fraction ratios between the LE and CT1 $\frac{k_f^r(CT)}{k_f^r(LE)} = 0, 1\cdot10^{-5}, 1\cdot10^{-4}, 1\cdot10^{-3}, 1\cdot10^{-2}, 1\cdot10^{-1}, 1$.

**2.11 Slower CT dissociation** alongside the weak LE dissociation has been reported for low-offset systems [2]. We can implement this finding by modelling a variable dissociation rate coefficient, so that $k_{d,CT}$ "vanishes" below $10^{11}s^{-1}$ when $k_{r,LE}$ dominates over $k_{d,LE}$. For small and *negative offsets*, $\eta_{EL}$ rises, reducing the significance of the bump somewhat. Fig. S5a-d (dashed lines) show a better data fit to all other metrics as well.

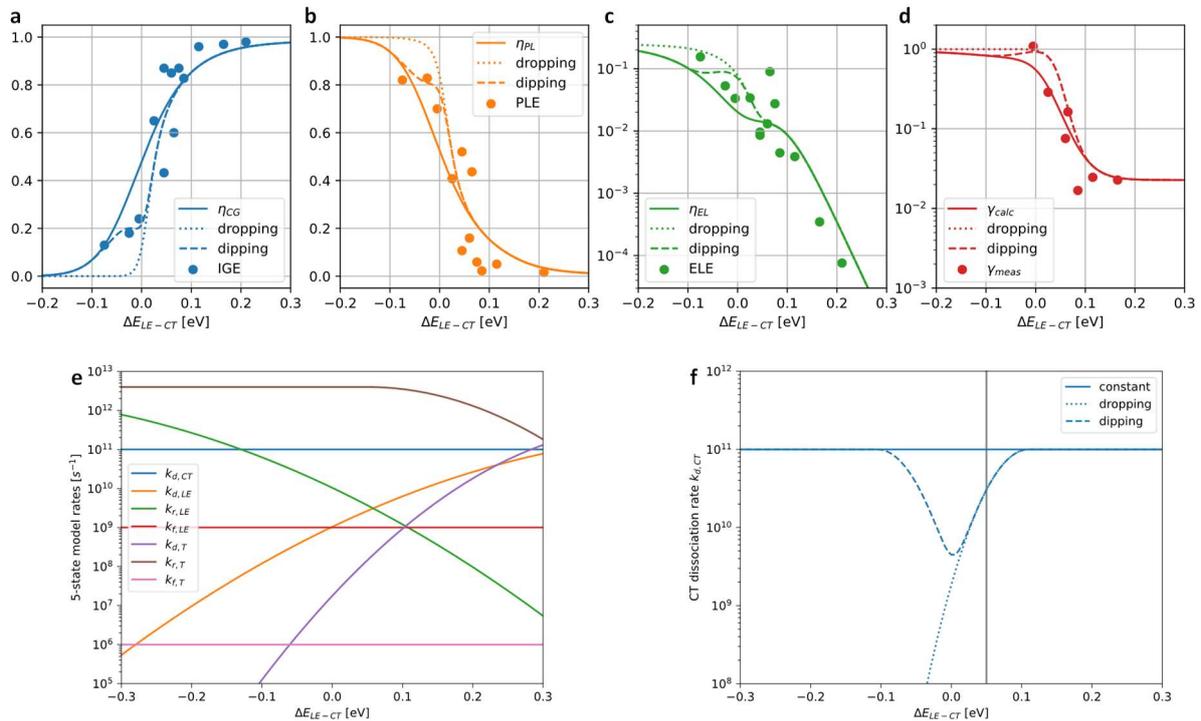

Figure S5: a-d) Similar to Fig. 4 in the main text. We compare the 5-state model with different CT dissociation rate behaviours: (solid) a constant CT dissociation rate coefficient $k_{d,CT} = 1\cdot10^{11}s^{-1}$ as in the main text; (dotted) the CT dissociation quickly drops below $1\cdot10^{11}s^{-1}$ when LE reformation dominates over LE dissociation, i.e. below $\Delta E_{LE-C} < 100 meV$; (dashed) the CT dissociation only dips by over an order of magnitude for small offsets $|\Delta E_{LE-CT}| < 100 meV$, in accordance with [2]. These three cases for $k_{d,CT}$ are shown in f). e) shows all rate coefficients used in the model, note the constant inverted regime for $k_{r,T}$.

## 3. Model result dependencies on input parameters

The effect of the different model parameters on the device metrics is illustrated here. Note that only a single parameter is changed per row, keeping all other parameters constant. Some parameters are closely linked, such as $\Delta E_{ST}$ and $k_{f,T}$, which have an almost identical impact on the metrics. A change in one can therefore be compensated by a change in the other.

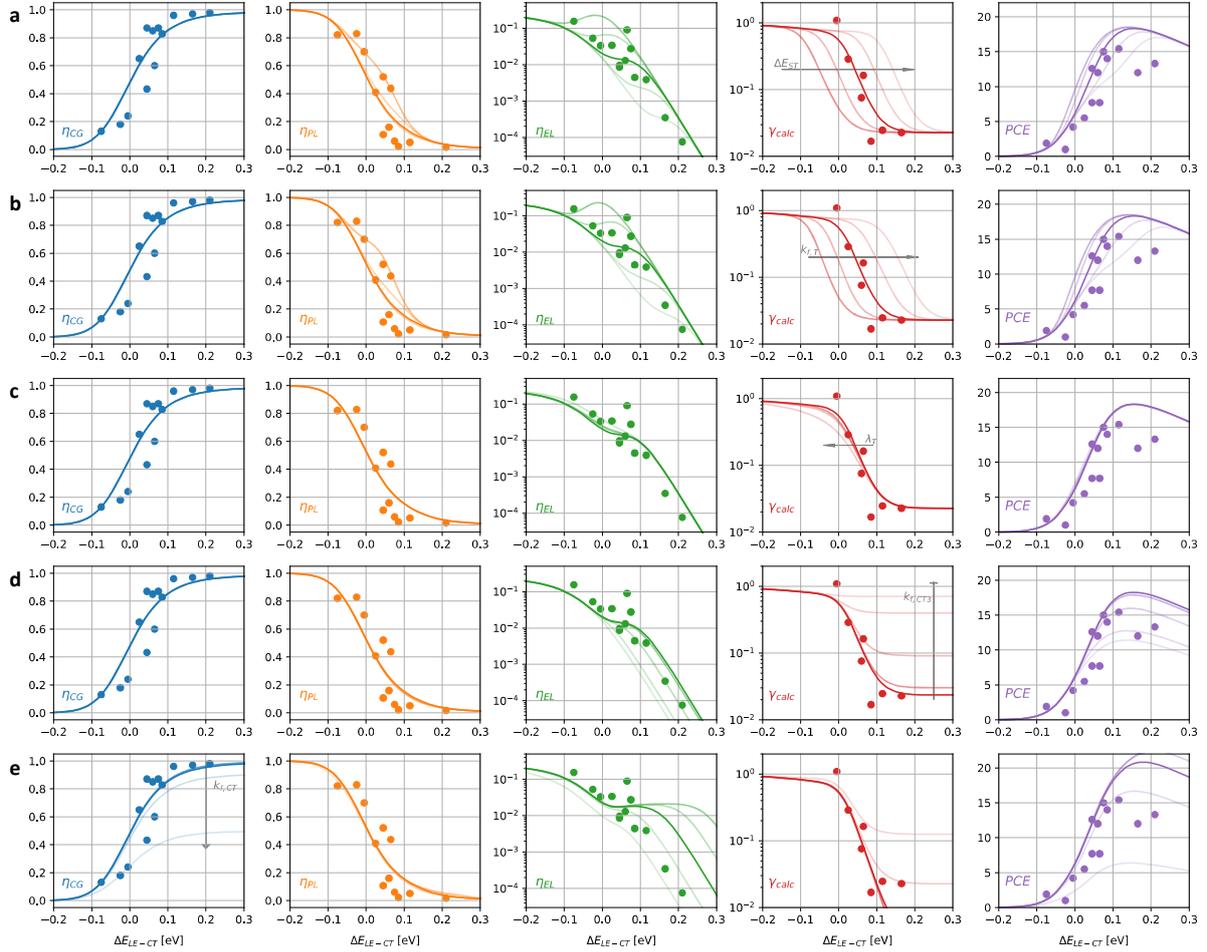

Figure S6: Changing different model parameters to reduce the ELQY-bump. An arrow in the plot indicates the changing parameter for the entire row and shows the direction of increase whereas the exact values are listed here, **bold** indicating the best fitting value. a) $\Delta E_{ST}[meV] = 200, 250, \mathbf{300}, 350, 400$ b) $log(k_{f,T} \cdot s) = 4, 5, \mathbf{6}, 7, 8$ c) $\lambda_T[eV] = \mathbf{0.2}, 0.4, 0.6, 0.8$ d) $log(k_{f,CT3} \cdot s) = 8, 9, \mathbf{10}, 11, 12$ e) $log(k_{f,CT} \cdot s) = 7, 8, \mathbf{9}, 10, 11$

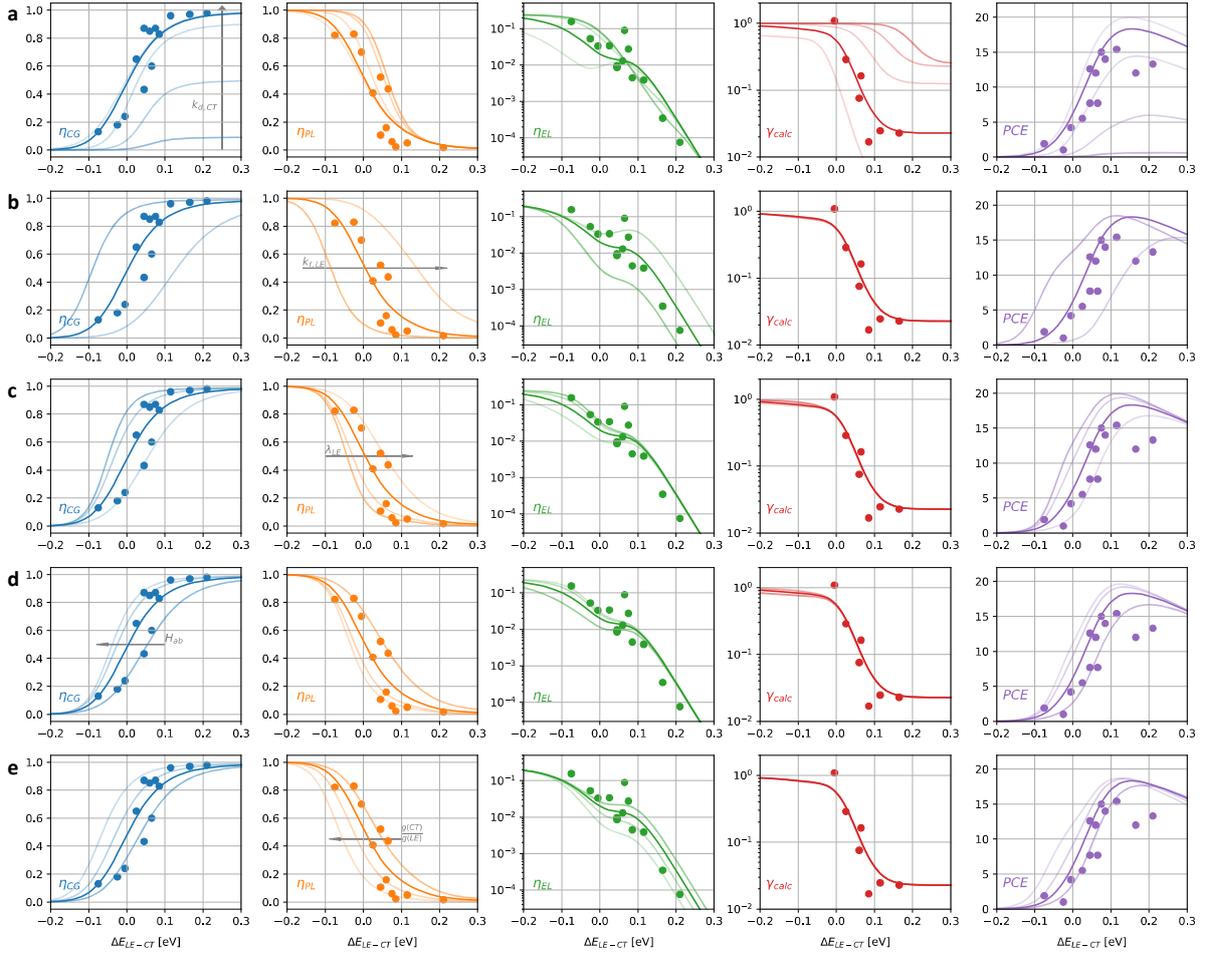

Figure S7: Continuation of Fig. S6. a) $log(k_{d,CT} \cdot s) = 8, 9, 10, \mathbf{11}, 12$  b) $log(k_{f,LE} \cdot s) = 8, \mathbf{9}, 10, 11, 12$  c) $\lambda_{LE}[eV] = 0.35, 0.45, \mathbf{0.55}, 0.65$  d) $H_{ab}[meV] = 5, \mathbf{10}, 15, 20$  e) $\frac{g(CT)}{g(LE)} = 5, \mathbf{10}, 15, 20$.

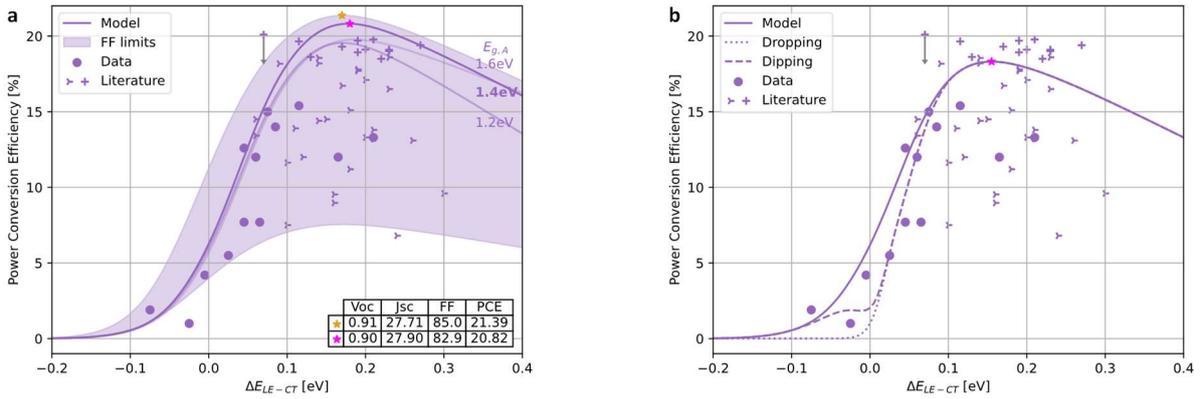

Figure S8: a) Same as Figure 6b but for the simple 5-state model, i.e. using the same CT decay rate coefficient $k_{f,CT} = 1 \cdot 10^9 s^{-1}$ for the generation and recombination pathways. PCE values are higher due to increased ELQY compared to if CS encounters to deeper CT states. b) Also same as Figure 6b but showing the difference if CT dissociation is constant (solid), dropping (dotted) or dipping (dashed), as described in Supplementary Note 2.11. The maximum PCE is unchanged. Shown only for acceptor bandgap of 1.4eV.

## 4. Other figures

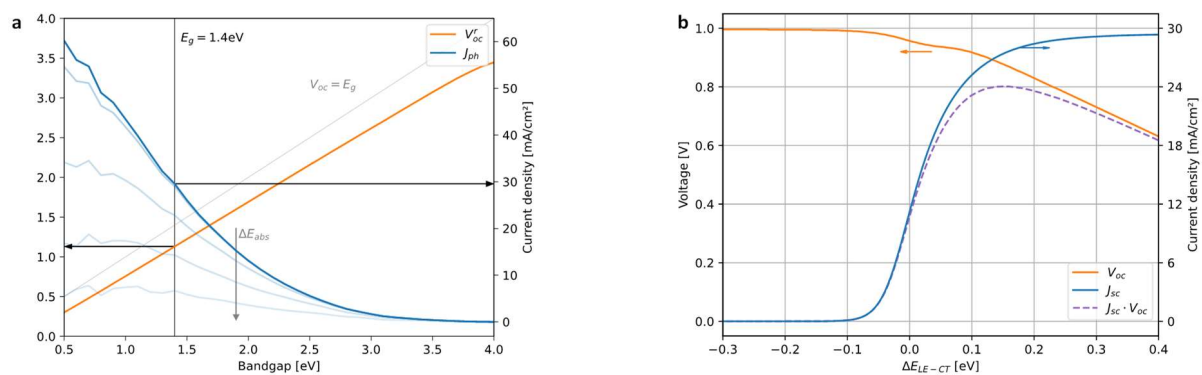

Figure S9: a) Radiative voltage limit (orange) and maximum photocurrent density (blue), assuming an absorption of 90% above the bandgap for AM1.5G. The vertical line shows $E_g = 1.4 eV$ and the arrows show the corresponding parameters ($V_{oc}^r = 1.130\, V$ and $J_{ph} = 29.555\, mA/cm^2$). The grey curve shows $V = E_g$. Lighter curves show the behaviour if the absorption bandwidth is reduced $\Delta E_{abs}/eV = \infty, 2, 1, 0.5, 0.2$ (indicated by the gray arrow). b) Open circuit voltage (orange) and short circuit current (blue) in the 5-state model (with CT modifications), showing a voltage loss of hundreds of meV compared to the radiative limit (at negative offset: $V_{oc}^r - V_{oc}^{max} = \Delta V^{min} = -\frac{k_B T}{q} ln\{0.02 * 0.25\} = 137$meV). An efficiency estimate is gained from multiplying both values (purple dotted), assuming the FF is unity.

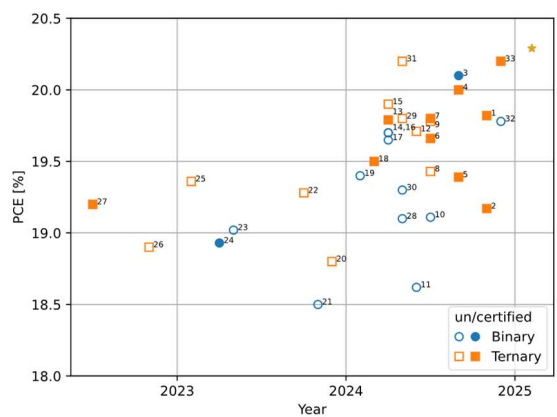

Figure S10: Record efficiencies of Organic Solar Cells made in the last 3 years: We differentiate between uncertified (open markers) and certified (filled markers) PCEs of binary (blue circles) and ternary (orange squares) architecture. References in Tab S3, note outlier no. 3. Our prediction of 20.29% is shown as a golden star.

# 5. Table of Definitions

| | | |
|---|---|---|
| Constants | | |
| | $k_B$ | Boltzmann constant |
| | $\hbar, h$ | (reduced) Planck constant |
| | $q$ | Elementary charge |
| | $\varepsilon_0, \varepsilon_r$ | Vacuum, relative permittivity |
| Solar Cell | | |
| | NFA | Non-fullerene acceptor |
| | D:A, PS:A | Donor:acceptor, Polystyrene:acceptor blend |
| | $V_{OC}$ | Open circuit voltage |
| | $J_{SC}$ | Short circuit current |
| | PCE | Power conversion efficiency |
| | FF | Fill factor |
| | $\Delta V_{nr}$ | Non-radiative voltage loss |
| | $\Delta_{HOMO}$ | HOMO level offset: donor-acceptor |
| | $E_g$ ($E_{g,A}$) | Bandgap energy (of the acceptor) |
| | $R$ ($R_{tot}, R_2, R_L$) | Recombination rate (total, bimolecular, Langevin) |
| | $\alpha$ | Figure of Merit |
| | $\mu_x, \mu_0$ | Mobility of charge carrier x, zero-field mobility |
| SQ calculation | | |
| | $\varphi_T^{bb}$ | Spectral photon flux density at black body temp. T |
| | $\Phi_{bb}$ | Photon flux density above the bandgap |
| | $J_{sat}, J_{ph}$ | Dark saturation current, Photocurrent (maximum) |
| | $V_{oc}^r$ | Radiative voltage limit |
| | $P_{inc}$ | Incident illumination power |
| | $\theta$ | Sun's solid angle from Earth |
| | $T$ ($T_c, T_S$) | Temperature (of the cell, Sun) |
| | $\Delta E_{abs}$ | Absorption band width |
| Model parameters | | |
| | $\Delta E_{ST}$ | Energy offset: singlet - triplet excitons |
| | $\Delta E_{LE-CT}$ | Energy offset: singlet exciton - charge transfer |
| | $H_{ab}$ | Transfer integral / Coupling strength of LE and CT |
| | LE, T | Localized exciton (singlet LE, triplet T1) |
| | CT = CT1, CT3 | Charge transfer state (singlet, triplet) |
| | CS | Charge separated ("free") state |
| | $G_{LE}$ | Generation rate into LE: photogeneration |
| | $G_{CS}$ | Generation rate into CS: electro-injection |
| | $k$ | Marcus rate coeff. (general) |
| | $k_{d,X}$ | Dissociation rate coeff. of state X |
| | $k_{f,X}$ | Decay rate coeff. of state X |
| | $k_{r,X}$ | Formation rate coeff. of state X |
| | $n_X$ | Population density of state X |
| | $\lambda$ ($\lambda_{LE}, \lambda_T$) | Reorganization energy (between LE-CT1, T-CT3) |
| | $\eta_x$ | Efficiency of process x, e.g. CT dissociation |
| | $\gamma$ | Langevin reduction factor |
| | $k_f^r(X)$ | Radiative fraction of decay of states X |
| | $g(X)$ | Density of states of state X |